\documentclass{aa}

\def \dj{d\kern-0.4em\char"16\kern-0.1em}
\def \Dj{\mbox{\raise0.3ex\hbox{-}\kern-0.4em D}}
\newcommand{\ind}[2]{#1_\mathrm{#2}}
\newcommand{\Rblr}[2]{\ind{R}{#1}^{\mathrm{#2}}}
\usepackage{graphicx}
\usepackage{epstopdf}
\usepackage{txfonts}
\usepackage{siunitx}
\usepackage{multirow}
\usepackage{amsfonts}
\usepackage{hyperref}
\hypersetup{
	    colorlinks,
	    linkcolor={red}, citecolor={blue}, urlcolor={magenta}
	    }
\usepackage{morefloats}
\usepackage{subfig}
\usepackage{etoolbox}
\makeatletter
\patchcmd\@combinedblfloats{\box\@outputbox}{\unvbox\@outputbox}{}{%
  \errmessage{\noexpand\@combinedblfloats could not be patched}%
}%
\makeatother

\DeclareSIUnit\solarmass{\ensuremath{M_\sun}}
\DeclareSIUnit\angstrom{\ensuremath{\AA}}
\begin{document}

   \title{Predicting the broad lines polarization emitted by supermassive binary black holes}
   \author{\Dj.\@ Savi\' c \inst{1,2}
          \and
          F.\@ Marin\inst{2}
          \and
          L.\@ \v C.\@ Popovi\' c\inst{1,3}
          }

  \institute{Astronomical Observatory Belgrade, Volgina 7, 11060 Belgrade, Serbia
	  \and
          Universit\'e de Strasbourg, CNRS, Observatoire Astronomique de Strasbourg, UMR 7550, 11 rue de l’Universit\'e, F-67000 Strasbourg, France
          \and
          Department of Astronomy, Faculty of Mathematics, University of Belgrade, Studentski trg 16, 11000 Belgrade, Serbia
          \and
          \email{djsavic@aob.rs}
          }
  \date{Received October 16, 2018; accepted December 16, 2018}

   \abstract
  {Some of Type-1 active galactic nuclei (AGNs) are showing extremely asymmetric Balmer lines with the broad peak redshifted or blueshifted by thousands of \SI{}{\kilo\meter\per\second}. These AGNs may be good candidates for supermassive binary black holes (SMBBHs). The complex line shapes can be very well due to the complex kinematics of the two broad line regions (BLRs). Therefore another methods should be applied to confirm the SMBBHs. One of them is spectropolarimetry.}
  {We rely on numerical modeling of the polarimetry of binary black holes systems since polarimetry is highly sensitive to geometry, in order to find specific influence of supermassive binary black hole (SMBBH) geometry and dynamics on polarized parameters across the broad line profiles. We apply our method to SMBBHs in which both components are assumed to be AGNs with distances at the sub-pc scale.}
  {We use a Monte Carlo radiative transfer code that simulates the geometry, dynamics and emission pattern of a binary system where two black holes are getting increasingly closer. Each gravitational well is accompanied by its own BLR and the whole system is surrounded by an accretion flow from the distant torus. We examine the emission line deformation and predict the associated polarization which could be observed.}
  {We model scattering induced broad line polarization for various BLR geometries with complex kinematics. We find that the presence of SMBBHs can produce complex polarization angle profiles $\varphi$ and strongly affect the polarized and unpolarized line profiles. Depending on the phase of the SMBBH, the resulting double-peaked emission lines either show red or blue peak dominance, or both the peak can have the same intensity. In some cases, the whole line profile appears as a single Gaussian line, hiding the true nature of the source.}
  {Our results suggest that future observation with the high resolution spectropolarimetry of optical broad emission lines could play an important role in detecting sub-pc SMBBHs.}
   \keywords{Galaxies: active galactic nuclei -- black holes -- polarization -- scattering}
   \maketitle
%
\section{Introduction}
According to the standard paradigm, every massive galaxy is expected to host a supermassive black hole (SMBH) in its center \citep{1995ARA&A..33..581K}. The typical mass range of those black holes is ranging between 10$^6$ and 10$^9$, with few examples of 10$^{10}$ solar masses cases \citep{2004ApJ...614..547S,2014MNRAS.442.2809W,2015ApJ...799..189Z}. The mass of the SMBH slowly evolves with time \citep{2009MNRAS.400.1451V} and is tightly correlated with the properties of the host galaxy it resides in (e.g., bulge mass,velocity dispersion, see \citealt{2013ARA&A..51..511K}). It is then crucial to better understand the evolution of SMBH in order to constrain galaxy formation models.
If accretion of matter from the surrounding environment is a natural way to increase the mass of the SMBH, it is a slow process that has difficulties to explain the most massive cases \citep{2010Natur.466.1082M}. In addition, only 60\% of the accreted mass is effectively transferred into the potential well, the rest being converted into high energy radiation \citep{2009PFR.....4...17D}. Another hypothesis for the evolution of SMBH is via mergers with other SMBHs \citep{2003ApJ...582..559V,2003ApJ...593..661V}. On large scales, dynamical friction is the main process that brings the SMBHs closer \citep{1980Natur.287..307B} but once the merging of the two host galaxies has been achieved, the final parsec problem onsets \citep{2003AIPC..686..201M}. Dynamical friction becomes inefficient when the two SMBHs form a bound binary; the system has no options to release energy and transfer angular momentum. One possible solution is that the spinning black holes lose energy by emitting gravitational waves \citep[GW,][]{1980Natur.287..307B}. The first discovery of GWs with frequency $\sim$\SI{d2}{\hertz} coming from stellar-mass binary BHs \citep{2016PhRvL.116f1102A} is a huge advancement in general relativity. The GW frequency for SMBBHs with mass range from \SI{d6}{} -- \SI{d9}{\solarmass} falls in the range from nanohertz to milihertz band and so far, none have been detected. In this frequency regime, pulsar timing arrays \citep[PTAs,][]{2015Sci...349.1522S} can be used for detecting GW by monitoring pulses from millisecond pulsars, however we are still waiting for the detection of such signatures that should be numerous. The occurrence of long-lived binary SMBHs signals appears to be too rare. Hence, are there really binary SMBHs?

Finding observational evidences of binary SMBHs is a difficult task. First of all, it is hard to spatially resolve at pc-scale the central part of the nearest galaxies with existing telescopes, therefore one has to find other methods to search for sub-pc SMBBHs. The emission of broad, double-peaked Balmer emission lines observed in the spectra of several active galactic nuclei (AGN) may (not) be associated with binary systems \citep{2003ApJ...599..886E,2009NewAR..53..133E}. During the merging effect of two galaxies, in  a sub-pc phase of SMBBH system, there is
enough gas which may produce an activity similar to the one observed in AGNs \citep{2012NewAR..56...74P}. Since AGNs have some comparable and well-known spectroscopic characteristics, one of the promising methods of the SMBBH detection is broadband spectroscopy, i.e. observations in a wide wavelength band including the emission lines \citep[see][for review]{2012NewAR..56...74P} can give some indications for SMBBH presence in the center of some active galaxies \citep[see e.g.][]{2012ApJ...759..118B,2015MNRAS.453.1562G,2016ApJ...822....4L}.

According to the standard theory, AGNs are powered by a supermassive black hole that releases tremendous amounts of energy throught accretion processes. A thermal continuum is arising from the accretion flow and line emission is dominated by emission from the so-called broad-line region (BLR) that surrounds the accretion disk. \citep{2008RMxAC..32....1G,2009NewAR..53..140G}. The BLR is a rotating, turbulent disc that is both optically and physically thick, and probably composed of numerous cloudlets of ionized gas. When this distribution of gas is seen face-on (i.e., from the AGN polar direction, which is free of opaque media) we see centrally-peaked line profiles. When the BLR is seen at a different inclination, a characteristic double-humped “disk-like” profile appears \citep{2003ApJ...599..886E}. However a significant fraction of AGNs show broad-line profiles that cannot be explained by this axisymmetric BLR model \citep[see, e.g.,][etc.]{1979ApJ...230..681C,1985PASP...97..734M,1990agn..conf...57N,2003A&AT...22..661G,2016ApJS..222...25S}. They show strong asymmetric displaced BLR peaks with the broad peak redshifted or blueshifted by thousands of km.s$^{-1}$. According to \citet{2009Natur.458...53B} those signatures could be due to a binary SMBH system, resembling a spectroscopic binary. As it was discussed by \citet{2012NewAR..56...74P}, the broad line profiles and their variability may indicate the SMBBH presence, however an additional evidence is needed to check it, as e.g. $\gamma$-ray and $X$-ray emission or polarization in the broad emission lines.

To test this hypothesis polarimetry is a natural tool since the geometry of the emitting and scattering system is expected to produce polarimetric features that are easily distinguishable from model to model \citep{2007A&A...465..129G,2012A&A...548A.121M,2014AdSpR..54.1341G}. A single SMBH surrounded by coplanar cylindrically-shaped scattering regions produces very low amounts of polarization when seen from a close to pole-on inclination \citep{2012A&A...548A.121M}. The polarization in the line shares similar values as the continuum and shows characteristic, wavelength-dependent
variations across the line profile\citep{2002MNRAS.335..773S,2014MNRAS.440..519A}. The polarization angle across the line profile for a single SMBH can indicate Keplerian-like motion, and consequently can be used for the black hole mass measurements \citep{2015ApJ...800L..35A,2018A&A...614A.120S}. The case of extremely asymmetric Balmer lines with large redshifted or blueshifted peaks could not be tested since the spectropolarimetric signal for binary SMBHs, each surrounded by its own BLR, is not known.

There is a number of publications which consider the broad line shapes of AGNs in the case of sub-pc SMBBHs \citep[see e.g.][etc.]{1983LIACo..24..473G,2000SerAJ.162....1P,2010ApJ...725..249S,2012ApJS..201...23E,2016Ap&SS.361...59S,2016ApJ...828...68N}, while the polarization effects in the line profiles was never considered in details. Exception is the observations \citep[][]{2010ApJ...717L.122R} and theoretical consideration \citep[][]{2017Ap&SS.362..110P} of the shift of polarized broad lines  for a kicked supermassive black hole. \cite{2010ApJ...717L.122R} gave an observational evidence that  quasar E1821+643\footnote{The quasar has highly shifted Balmer lines around 1000 km s$^{-1}$ and a red asymmetry \citep[see][]{2016ApJS..222...25S}} may be an example of gravitational recoil, i.e.\,they found that broad Balmer lines indicate the kick off velocity of  $\sim $2100 km s$^{-1}$ in polarized light. \cite{2017Ap&SS.362..110P} also considered recoiling black hole, taking that  kick radius is similar to the BLR dimension and found that polarization data in this case can give an estimation of the kick off velocity.

The purpose of our study is to explore, for the first time, the polarization parameters across the broad lines in the case of an emission by a sub-pc scale SMBBH system. By doing so, we aim to predict what should be the observational signature we expect from those yet-to-be-confirmed sources. We consider a model of sub-pc supermassive binary black holes, where each of the BH components has own accretion disk and BLR. We consider equatorial scattering of such complex system on the inner part of the torus, and we modeled the Stokes parameters which can be observed from the system. The paper is organized as followed: In Section \ref{S.model} we describe the model and the basis parameters of the model which we used to calculate the polarization parameters. In Section \ref{S.results} we present and analyze obtained results of our simulations, where we take different parameters of SMBBHs. Finally, in Section \ref{S.discussion} we discuss our results and in Section \ref{S.conclusions} we outline our conclusion.

\section{Model setup} \label{S.model}
We model SMBBH system as two black holes orbiting around the common center of mass under the force of gravity. This is a well known problem for which it was shown that it is equivalent to the problem of a single body with reduced mass $\mu$ moving in an external gravitational field \citep{landau1969mechanics} which is determined by the total mass of the system:

\begin{equation}
 M = M_1 + M_2,
\end{equation}
where $M$ is the total mass, and $M_1$ and $M_2$ are masses of each component. The reduced mass $\mu$ is

\begin{equation}
 \mu = \frac{M_1M_2}{M}.
\end{equation}
In general, the body $\mu$ moves in elliptical trajectory with semi-major axis $a$ and eccentricity $e$. The relationship between the orbital period $P$, orbital frequency $\Omega$, $M$ and $a$ is given by the Kepler's third law:

\begin{equation}
 \Omega=\frac{2\pi}{P}=\sqrt{\frac{GM}{a^3}},
\end{equation}
where $G$ is gravitational constant. This relation is valid for any eccentricity $e$. Each component is moving around the center of mass in elliptical orbit with the same eccentricity $e$. Both ellipses lie in the same plane and have one common focus. The semi-major axes are inversely proportional to the masses:

\begin{equation}
 \frac{a_1}{a_2} = \frac{M_2}{M_1},
\end{equation}
and they satisfy the equation:

\begin{equation}
 a = a_1 + a_2.
\end{equation}
Our goal is to create a simple, yet comprehensive model, without introducing hydrodynamic simulations and three body problem solving. A second model, based on hydrodynamic simulations is presented in Section \ref{s.spirala}. In this work, we are considering the case with $e = 0$, i.e.\ orbits are circular. and with both black holes having the same mass $M_1 = M_2 = \SI{5d7}{\solarmass}$, i.e.\ the mass ratio $q = M_2/M_1 = 1$.

We have made two assumptions in our model: one is that both SMBHs have accretion disks and the corresponding BLRs and the second is that both the accretion disks and the scattering region are coplanar.  We expect to have near coplanar accretion disks and scattering region (torus) because of following reason: In gas rich mergers, where the evolution of the SMBBHs is driven by interaction with the surrounding gas, the accretion onto the black holes leads to the alignment of black holes spin with the angular momentum of the binary \citep{2007ApJ...661L.147B} which effectively lowers the kick velocity \citep{2010MNRAS.402..682D}. The timescale of the angular momentum aligning with the individual spin of each component is few hundreds of times shorter than the timescale for which the angular momentum of the binary aligns with the angular momentum of the inspiraling circumbinary gas, unless the mass ratio is extreme \citep{2013ApJ...774...43M}. In case that the accretion occurs in the opposite direction of the binary rotation, there will be a misalignment of various axes on a timescale of the order of a fraction of the whole binary evolution time. As was mentioned above, each black hole has an accretion disc surrounding it, from which the isotropic continuum radiation is emitted. We used point source approximation for disc emission with emissivity given by a power law: $F_C \propto \nu^{-\alpha}$ where $\alpha$ is spectral index equal to 2. Both black holes are surrounded by the BLR. Depending on the distance between the black holes, we treated four different SMBBH cases: \textbf{distant}, \textbf{contact}, \textbf{mixed} and \textbf{spiral}. We modeled BLR with flared-disk geometry \citep{2007A&A...465..129G} with half-opening angle of \ang{25} which gives a covering factor of the order of 0.1 \citep{2013peag.book.....N}. The size of the BLRs were set to few light days with BLR inner radius $\Rblr{in}{BLR} = 3$ and BLR outer radius $\Rblr{out}{BLR} = 12$ light days. These values for BLR inner and outer radius were chosen to reproduce typical BLR velocity values of few thousands \SI{}{\kilo\meter\per\second} \citep{2004ApJ...613..682P,2005ApJ...629...61K}. This was done for all cases except for the spiral one.

\begin{figure}
   \centering
   \includegraphics[width=\hsize]{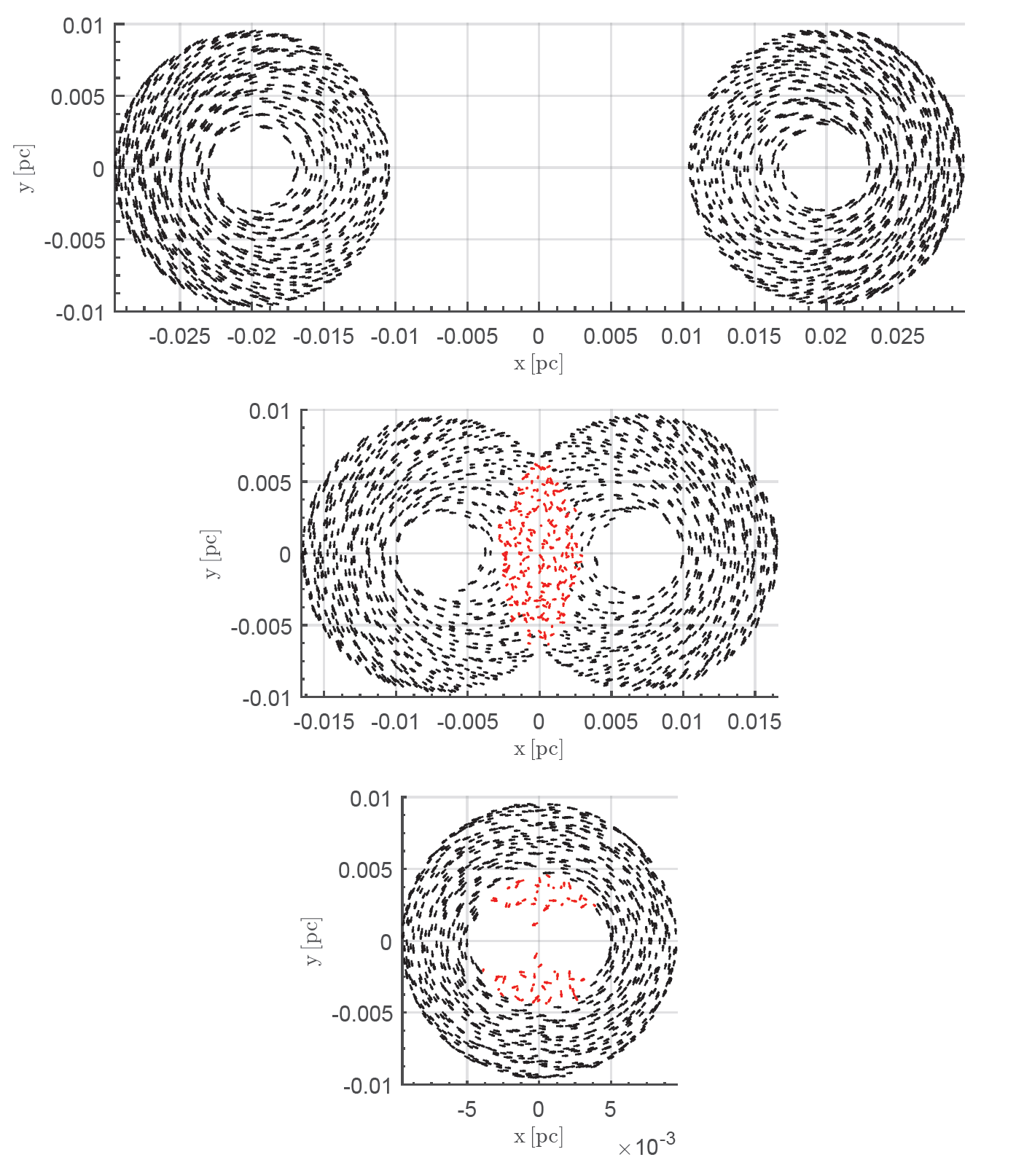}
      \caption{Geometry and kinematics of the SMBBH for each model. From top to bottom: Distant, Contact, and Mixed. Black arrows show the non-perturbed velocity field, while red ones are for clumps with additional random component of the velocity.}
      \label{f:model}
\end{figure}

\textbf{Distant:} Both BLRs are distinctive and each black hole affects only the dynamics of the BLR it is surrounded with. Each BLR cloud has two velocity components: Keplerian motion around the black hole plus additional motion due to the binaries orbiting each other (see Fig.\,\ref{f:model}, top panel). Black holes are at the orbital distance $a = 47.6$ light days which corresponds to the orbital period of approximately 75 years.

\begin{figure}
   \centering
   \includegraphics[width=\hsize]{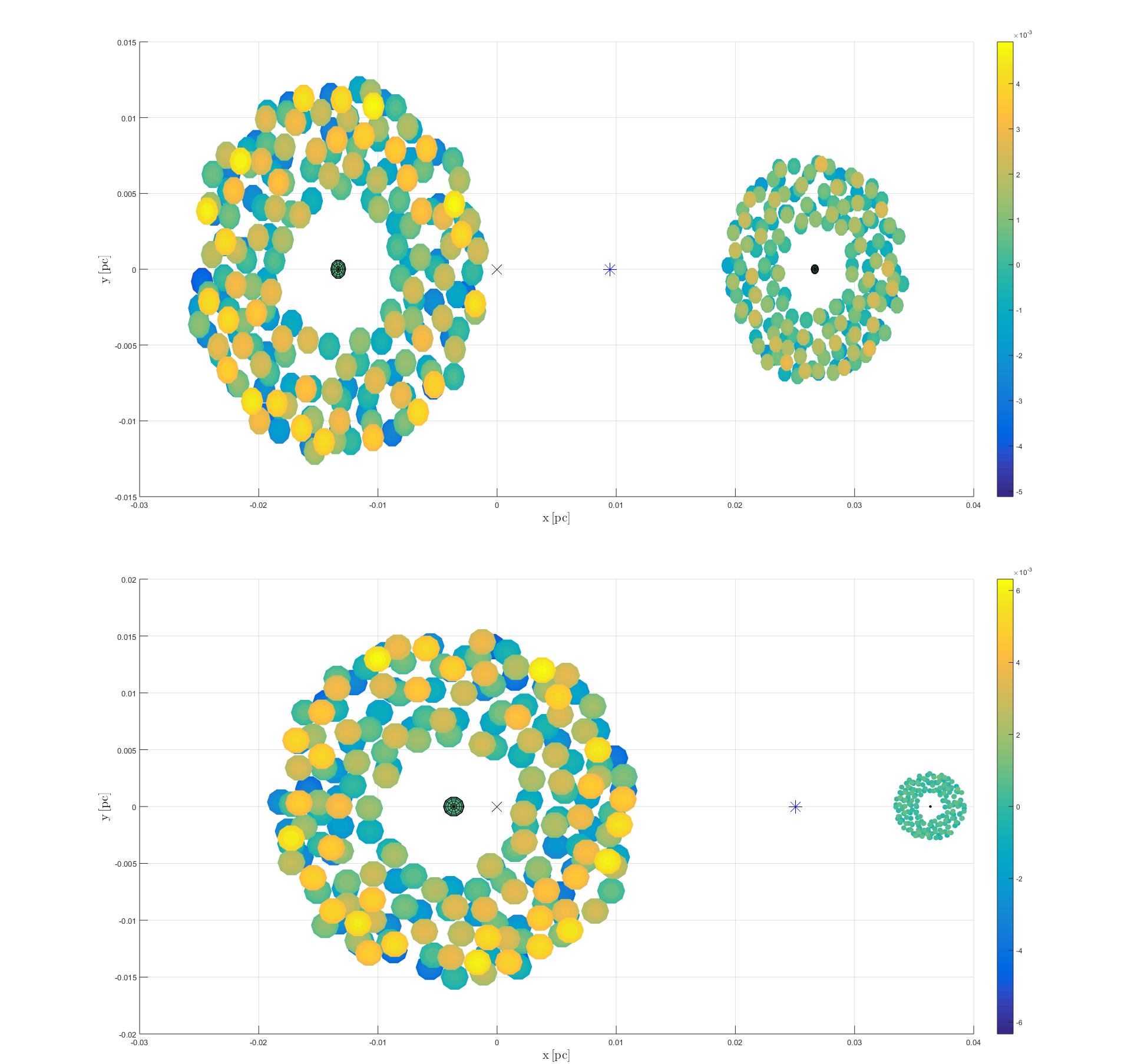}
      \caption{Distant model with mass ratio $q = 0.5$ (upper panel) and with $q = 0.1$ (lower panel). Black cross is showing the center of mass, while the blue asterisk symbol marks the L1 Lagrangian point. Color bar is denoting the vertical offset from the $xy$ plane.}
         \label{f:q0501}
\end{figure}

Additionally, we simulated two models with mass ratio $q = 0.5$ and $q = 0.1$ for this case. Assuming that photoionization and recombination following radiative de-excitation is the main mechanism for the emission of broad Balmer lines, the BLR size scales with luminosity in the form of $\ind{R}{BLR} \propto L^{0.5}$ \citep{2005ApJ...629...61K}. We used mass luminosity relation $\ind{M}{BH}\propto L^{0.7}$ \citep{2002ApJ...579..530W} in order to obtain the BLR size depending solely on mass of each component. An illustration for these two cases is shown in Fig.\, \ref{f:q0501}.

\textbf{Contact:} Black holes are separated by $a = 16.7$ light days with orbital period of 15.5 years, which allows for certain parts of the BLRs to overlap (Fig.\,\ref{f:model}, middle panel). In this regime, the BLR kinematics is similar as in the previous model, except for the overlapping part where we assigned chaotic component to the velocity for each clump due to chocks, stirring and inelastic collisions.

\textbf{Mixed:} For this model, black holes are much closer to each other, at the orbital separation of 3 light days and with orbital period of 1.2 years. On Fig.\,\ref{f:model}, third panel, clumps denoted in red are the ones with additional chaotic component, while for the rest we calculated velocity as if in the center was a single SMBH with mass equal to the sum of binary components.

\begin{figure}
   \centering
   \includegraphics[width=\hsize]{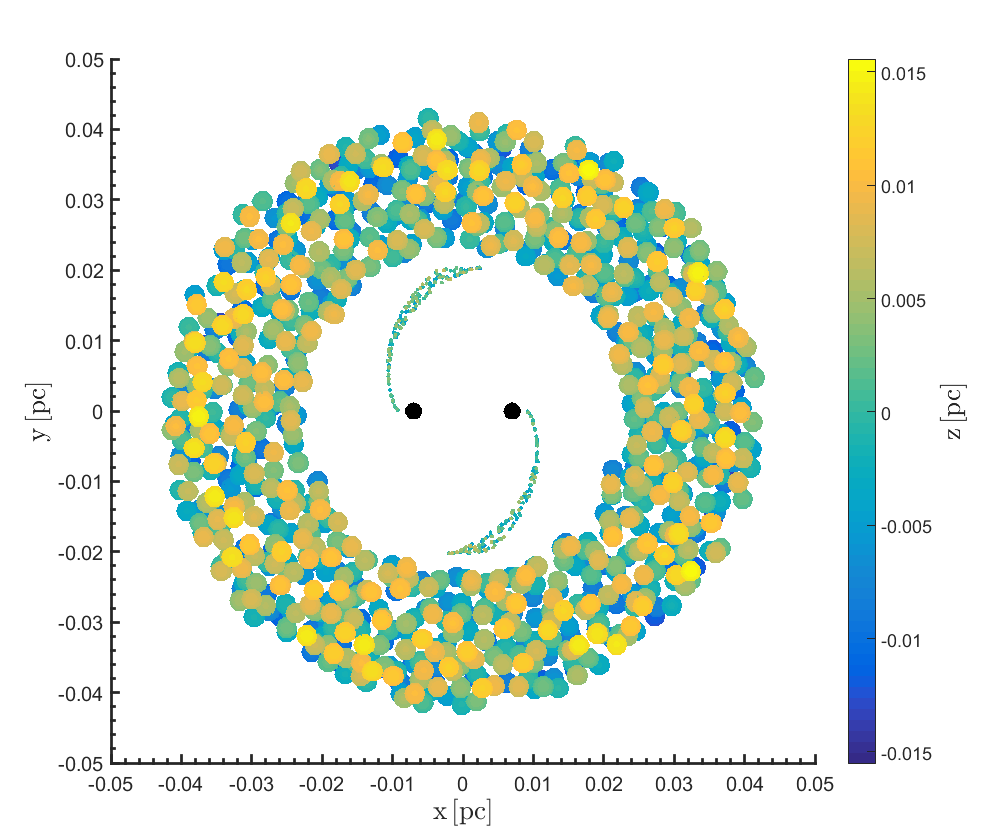}
      \caption{SMBBHs (black circles) with spiral arms surrounded by a CB disk. Each spiral is modeled with 500 clumps. The CB is modeled with 1000 clumps. Color bar is denoting the vertical offset from the $xy$ plane.}
         \label{f:spiral}
\end{figure}

\textbf{Spiral:} Hydrodynamic simulations involving subparsec SMBBHs have shown that black holes are surrounded by a common circumbinary (CB) disc. Accreting gas around the binaries forms a low density cavity inside the CB disc \citep{2008ApJ...672...83M,2009MNRAS.393.1423C}. It was found that the accretion streams are in the form of spiral arms with higher density that is connecting mini accretion disk of each black hole with the surrounding CB disk \citep{2012ApJ...755...51N,2015ApJ...807..131S}. In this scenario, the cavity is of the order of $a$, and the CB disk extends from $1.5a$ to $3a$. Following the similar setup as \citet{2015JApA...36..513S}, we built a SMBBH model with spiral arms and the surrounding CB disk in order to investigate the polarization signatures coming from the SMBBH. We keep the same mass of each component to be \SI{5d7}{\solarmass} with the orbital separation the same as in the case for contact model $a = 16.7$ light days. We approximated spiral arms with logarithmic spirals with boundaries in polar coordinates given as:

$$R_1 = \frac{a}{2}e^{b\phi} < R(\phi) < R_2 = \frac{a}{2}e^{B\phi},$$

\noindent where $b$ and $B$ are parameters describing the wrapping of the spirals. We chose wrapping parameters to be $B = 0.55$ and $b = 0.45$. This set of parameters for $b$ and $B$ were chosen in order to have two distinct spirals with single winding and to avoid mixture or interaction of the spirals. We chose the half-opening angle for the spirals and the CB disk to be \ang{20}. An illustration of the model is shown at Fig.\,\ref{f:spiral}. For kinematics of the spirals, we used the rotation of absolute rigid body, i.e.\, the spirals are stationary in the rotating reference frame of the SMBBH. The CB is under the Keplerian motion around the common center of mass. The system is again surrounded by the same scattering region as in previous models, with the same radial optical depth in the equatorial plane.

The BLR is represented by thousands of clumps. The volume filling factor of the BLR of 0.25, as constrained from simulations and observations \citep{2015A&A...577A..66M}. Total number of clumps per model as well as the other parameters used in the model are listed in Table \ref{t:model}

\subsection{The scattering region}
\begin{figure}
   \centering
   \includegraphics[width=\hsize]{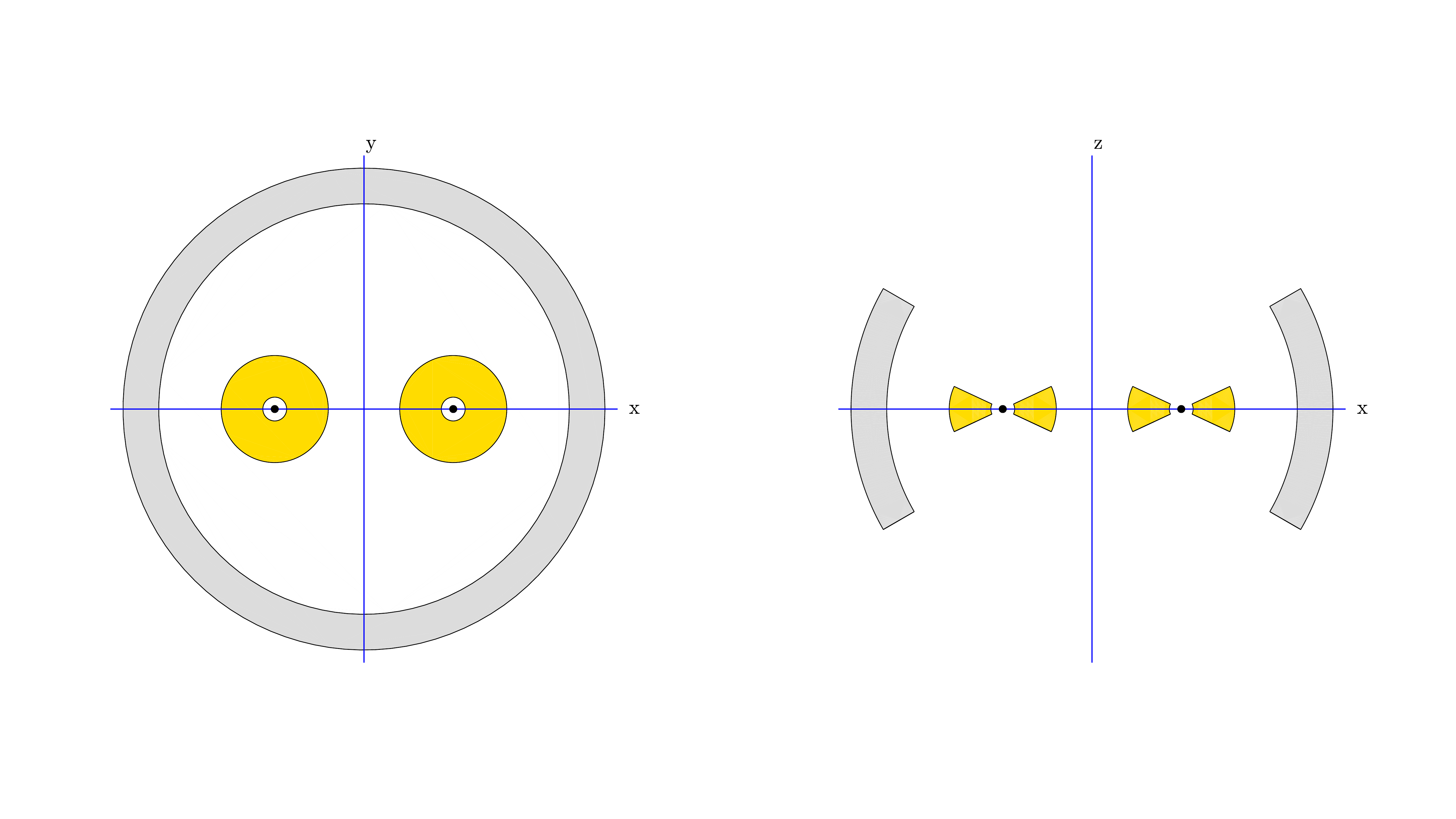}
      \caption{Cartoon illustrating equatorial scattering region. Left figure shows the face-on view, while on the right the same geometry is shown when viewed edge-on. An example is shown for the case with the two BLRs being separated. The BLRs are shown in yellow. Scattering region is denoted in grey. }
      \label{f:model_sr}
\end{figure}

Optical continuum and line polarization properties typically found in Type-1 objects can be produced by electron scattering of a flattened distribution that is surrounding the accretion disk and the BLR \citep{1984ApJ...278..499A,2005MNRAS.359..846S}.
The scattering region is modeled with flared-disk geometry with inner and outer radius of 0.1 and 0.5 parsec. The half-opening angle is \ang{30} with respect to the equatorial plane. Electron concentration is chosen in such a way that the total radial optical depth in the equatorial plane for Thomson scattering is 3, which is enough to produce typical degree of polarization that is found in Type-1 objects \citep{2012A&A...548A.121M}. An illustration of the scattering region surrounding the central engine is illustrated on Fig. \ref{f:model_sr}

\begin{table*}
\caption{Description of the 3 SMBBH model. $V_1$ and $V_2$ are orbital velocities and $q$ is the mass ratio.}
\centering
\begin{tabular}{lcccccc}
\hline\hline
Model & Orbital separation $a$ & Orbital period $P$ & Number & $V_1$ & $V_2$ & q\\
	  & \SI{}{light\, days}      & \SI{}{years}	       &  of clouds & \SI{}{\kilo\meter\per\second} & \SI{}{\kilo\meter\per\second} &\\
\hline
Distant & 47.65 & 75.0 & 2000 & 1639 & 1639 & 1.0\\
Distant & 47.65 & 75.0 & 2000 & 1093 & 2186 & 0.5\\
Distant & 47.65 & 75.0 & 2000 & 298 & 2980 & 0.1\\
Contact & 16.68 & 15.5 & 1600 & 2771 & 2771 & 1.0\\
Mixed   & 2.978 & 1.2  & 1000 & 6558 & 6558 & 1.0\\
\hline
\end{tabular}
\label{t:model}
\end{table*}

\subsection{Numerical simulations}

Assuming that AGN polarization arises predominantly from scattering in non-jetted systems, we apply full 3D radiative transfer with polarization using the publicly available code \textsc{stokes} \citep{2007A&A...465..129G,2012A&A...548A.121M,2015A&A...577A..66M,2018A&A...615A.171M,2018A&A...611A..39R}. It is based on Monte Carlo algorithm, for which a vast literature already exist, and with 3D kinematics fully implemented in spherical coordinates. The code follows the trajectory of each photon through the model space, from their creation, until they are being registered by the web of virtual detectors positioned all over the sky. The net Stokes parameters $I$, $Q$, $U$ and $V$ are thus being determined and other physical quantities may be inferred, namely degree of linear polarization (PO), polarized flux (PF) and polarization position angle ($\varphi$). Originally, the code was developed for studying optical and UV scattering induced continuum polarization in the radio-quiet AGNs, but nowadays it is widely used for studying polarization of many astrophysical phenomena \citep{2014sf2a.conf..103M}. We used the intermediate 2.04 version of the code \textsc{stokes} which is not yet publicly available\footnote{http://www.stokes-program.info/}. We adopt the same convention as \citet{2007A&A...465..129G}: we defined $\varphi$ to be \ang{0} when the polarization angle is perpendicular to the projected symmetry axis of the model. When $\varphi$ is \ang{90}, the polarization angle is parallel to the symmetry axis of the model.

\section{Results} \label{S.results}

\begin{figure*}
   \centering
   \includegraphics[width=\hsize]{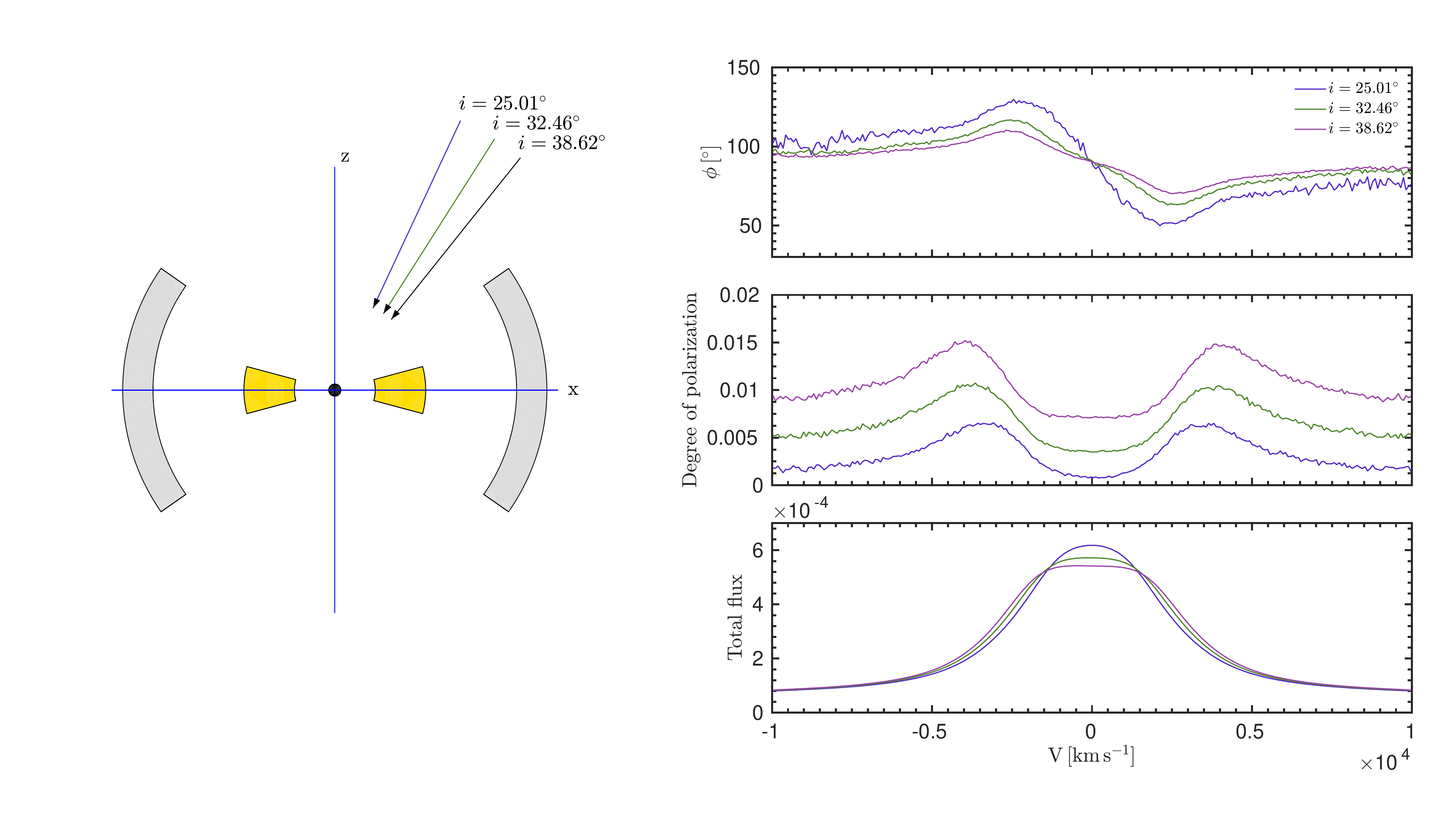}
      \caption{On the left panel, an illustration of the model with a single SMBH in the center surrounded by a BLR (yellow) and the scattering region (gray) is shown. Right panel: the profiles of polarization angle (top), the degree  polarization (middle), total flux (bottom) when viewed from two intermediate inclinations. Polarization angle is given in degrees ($^\circ$). We point out that the degree of polarization is given as fraction units and is lower than the ones we obtain in the following section due to the different size and optical depth of the scattering region. The total flux is given in arbitrary units. Model parameters are the same as the ones given by \citet{2018A&A...614A.120S}.}
         \label{f:single}
\end{figure*}

We simulated the different SMBBH scenarios presented in the previous section with different kinematics of the BLR depending on the model. In the following, we thoroughly investigate the results for each case. For clarity and easy comparison, we present the results of a model with a single SMBH in the center with mass $\ind{M}{bh} = \SI{d8}{\solarmass}$, so the reader could have a clearer picture when comparing the results for a single SMBH scenario with a SMBBH. The result of the single SMBH model is given in Fig.\,\ref{f:single}, the results for a SMBBH with the same center of mass in Fig. 6 and the numerous results for all the SMBBH scenarios are shown in Appendix.

For Type-1 objects, for a single case scenario i.\,e.\,a single black hole and a single BLR surrounded by a dusty torus, in the case for equatorial scattering, the $\varphi$ shows symmetric swing around the line center \citep{2005MNRAS.359..846S,2014MNRAS.440..519A,2015ApJ...800L..35A,2018A&A...614A.120S}. This feature was very well observed in few objects (e.g.\,Mrk 6, NGC 4051, NGC 4151) and can be used for measuring masses of SMBHs using polarization of broad line profiles \citep{2014MNRAS.440..519A,2015ApJ...800L..35A,2018A&A...614A.120S}.

\subsection{Distant}

\begin{figure*}[httb]
\centering
    \includegraphics[width=9cm]{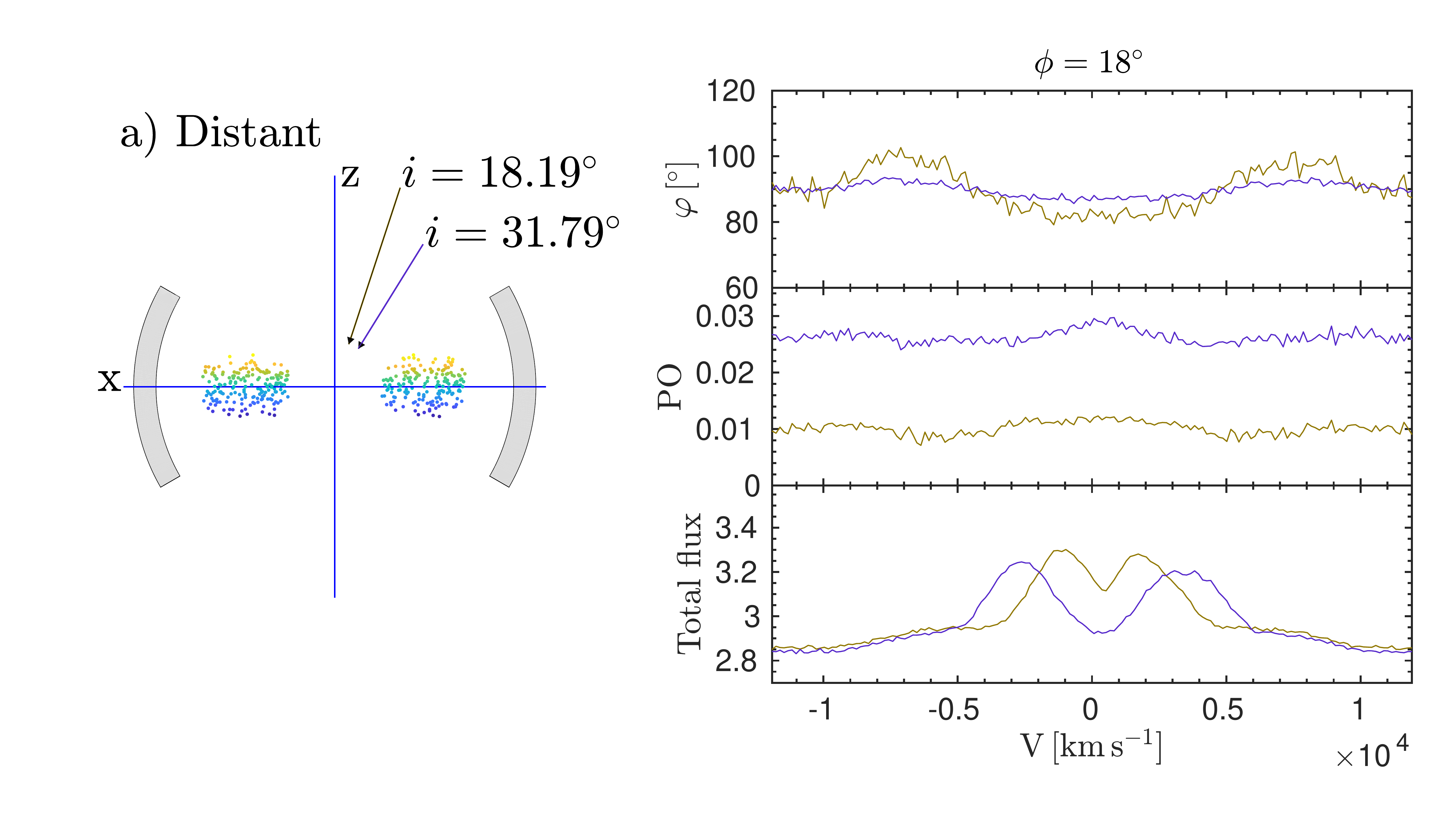}
    \includegraphics[width=9cm]{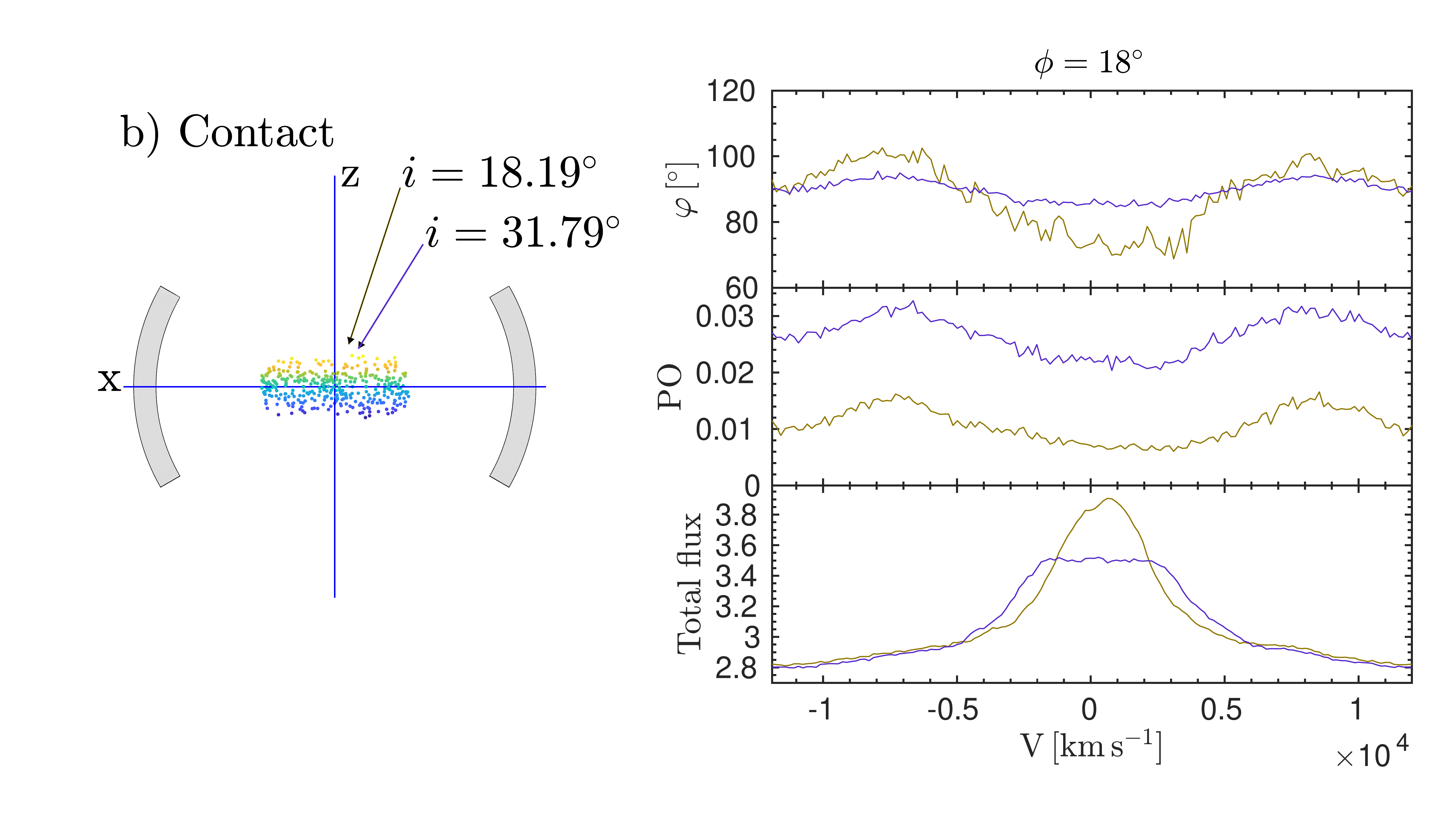}\\
    \includegraphics[width=9cm]{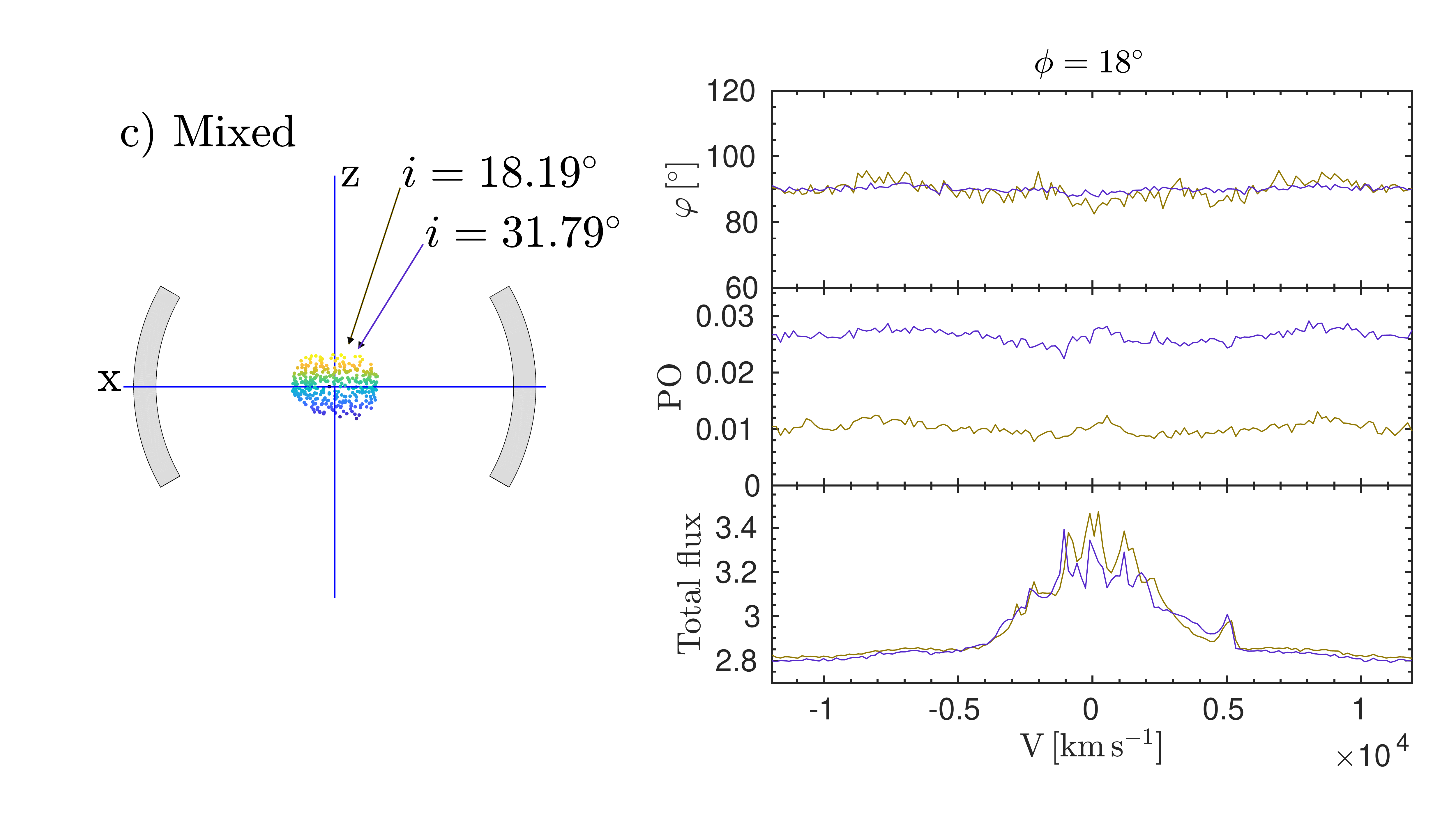}
    \includegraphics[width=9cm]{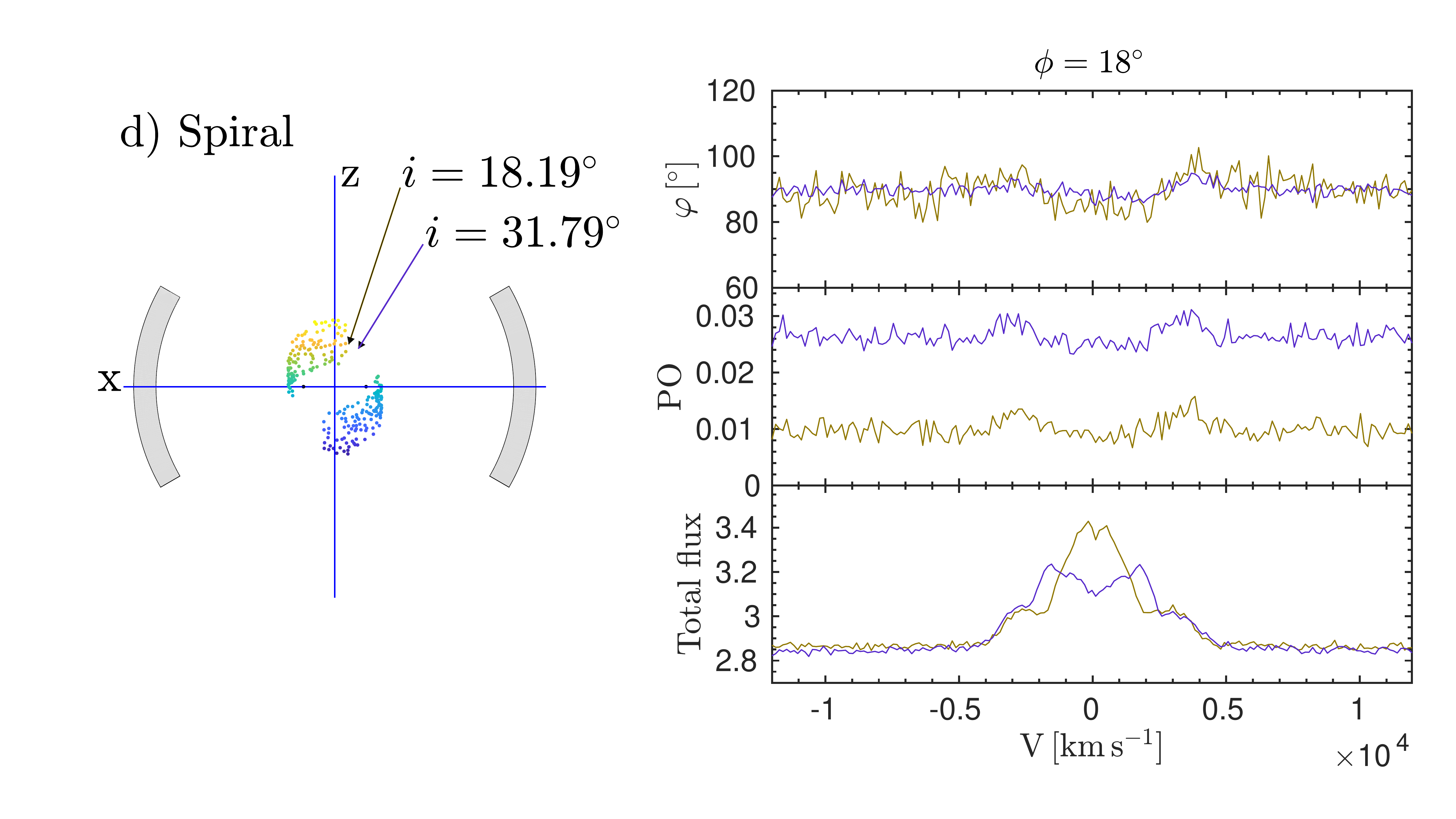}\\
  \caption{On the left panels the illustration of each model with SMBBH in the center: Distant (a), Contact (b), Mixed (c), Spiral (d). On the right panels, from top to bottom are $\varphi$, PO and TF for two viewing inclination and for azimuthal viewing angle $\phi = \ang{18}$.}
  \label{f:all}
\end{figure*}

In Figs.\,\ref{f:all} (panel a) and \ref{Distant_ES12_PA} we show the simulated $\varphi$-profiles for two viewing inclinations $i$ and for different azimuthal viewing angles $\phi$. We can see that profiles of $\varphi$ are complex and differ much from the profiles for the single black hole scenario. For a fixed viewing $\phi$ the $\varphi$-profiles show similar profiles with the peaks most prominent when viewed towards face-on inclinations. For different azimuthal viewing angles, $\varphi$-profiles are quite diverse. This diversity is the result of different velocity projections towards the observer since the model is not azimuthally symmetric. The $\varphi$-profiles are symmetric with respect to the line center which is not the case for a single case scenario where the swing occurs.

Typical degree of polarization PO found for Type-1 objects is around 1\% or less. Our simulations show that PO is in the range between 1\% and 4\% (Fig.\,\ref{Distant_ES12_PO}). This unusually high PO is due to the high radial optical depth of the scattering region. It is inclination dependent and it is increasing when observing from face-on towards edge-on viewing inclinations as expected from Thomson law. For some $\phi$ (Fig.\,\ref{Distant_ES12_PO}, top left and bottom right panels), PO profile peaks in the line wings and has a minimum value in the line core. This is the same as in the case for a single black hole scenario and it was confirmed observationally \citep[e.g.\, Mrk 6][]{2002MNRAS.335..773S,2014MNRAS.440..519A}. However, this is not the case for all $\phi$ and we can see the opposite situation -- PO peaks in the line core and has minimum in the line wings.

The total flux shows variability in the line profiles \ref{Distant_ES12_TF}. Line profiles are sensitive both to viewing inclinations and viewing azimuthal angles. In general, double-peaked profiles are observed, with line width being broader when observing from face-on towards edge-on inclinations. Line widths are different with respect to $\phi$ with the broadest lines coming from the direction when $\phi = \ang{90}$ or $\phi = \ang{270}$ (Fig. \ref{Distant_ES12_TF}, middle upper and bottom panels). Some viewing angles are showing single-peak lines (Fig.\,\ref{Distant_ES12_TF}, top left and bottom right panels) and the corresponding PO profiles are as in the case for a single black hole scenario. This means that in the certain phase, we would not be able to observationally distinguish between the SMBBHs and SMBHs from the unpolarized optical spectra. However for this case, $\varphi$ is showing different profile than expected, which could provide more insight if the SMBBHs is situated in the center.

For Distant model with mass ratio $q = 0.5$ we can expect asymmetric profiles for $\varphi$, PO and TF (Figs.\, \ref{Distant_ES12_q05_PA}, \ref{Distant_ES12_q05_PO}, \ref{Distant_ES12_q05_TF}). The $\varphi$ is having similar profiles as for the case with mass ratio $q = 1$ except that peaks are not symmetric and they have different intensities. When compared with the previous case, the $\varphi$-profile is similar except for azimuthal viewing angles $\phi = \ang{224}$ where the profile is flat in the core (Fig.\,\ref{Distant_ES12_q05_PA}, lower left panel), or an additional swing can be noticed in the core for $\phi = \ang{342}$ (Fig.\,\ref{Distant_ES12_q05_PA}, upper right panel).

Degree of polarization is having profiles with the same shape as for the previous case except that they are asymmetric and it is the case for all viewing angles. We obtained the same order of polarization with the same inclination dependency (Fig.\,\ref{Distant_ES12_q05_PO}).

The unpolarized line is showing a displaced single peak profiles when viewed almost face-on for most of the azimuthal viewing angles, except when $\phi = \ang{224}$ and $\ang{270}$ where a clear double-peaked profile can be observed (Fig.\,\ref{Distant_ES12_q05_TF}, bottom left and middle panels). For intermediate inclinations, line profiles are asymmetric with double peaks and with different line shifts depending on the azimuthal viewing angles (Fig.\,\ref{Distant_ES12_q05_TF}).

For the same model with $q = 0.1$, we obtained similar profiles as before for $\varphi$, PO and TF (Figs.\, \ref{Distant_ES12_q01_PA}, \ref{Distant_ES12_q01_PO}, \ref{Distant_ES12_q01_TF}), however they are more asymmetric than for the case with $q = 0.5$. The $\varphi$ is having similar profiles as for the cases with $q = 1$ and $q = 0.5$ with asymmetry highlighted (Fig.\,\ref{Distant_ES12_q01_PA}).

The degree of polarization is showing profiles with the same shape in the same way as before for all viewing angles. Polarization is of the same order with the same inclination dependency (Fig.\,\ref{Distant_ES12_q01_PO}).

The unpolarized flux is showing complex asymmetric profiles with line peaks having different positions as the system is viewed in different orbital phases (\ref{Distant_ES12_q01_TF}). When $q = 0.1$, the more massive component is having smaller orbital velocity and it is much smaller compared to the Keplerian velocity of the BLR clouds surrounding it. For the less massive component, orbital velocity is of the same order in comparison with the Keplerian velocity of the BLR clouds surrounding it, which contributes to higher line shift. With these two effects combined, we observe highly asymmetric line profiles which significantly vary with the orbital phase.

\subsection{Contact}
This scenario is geometrically similar with the previous one with the SMBBHs being closer and allowing additional chaotic velocity component will affect the line profile mostly around its core. Simulated $\varphi$ is shown on Figs.\,\ref{f:all} (panel b);\ref{Contact_ES1_PA}. The $\varphi$ profiles are also similar
as in the case for separated BLRs. Figures \ref{Contact_ES1_PA} (left panels; upper and middle right) clearly show two minima in the wings and a maximum in the line core; or minimum in the line core and maximum in the line wings. The observed $\varphi$ profile is the most sensitive to random velocity when the system is viewed from $\phi = \ang{90}$ and $\phi = \ang{270}$ (Fig. \ref{Contact_ES1_PA}, middle up and bottom panels), for which we observe two minima and almost flat profile in the core. For $\phi = \ang{342}$ (Fig. \ref{Contact_ES1_PA}, bottom right panel) we see one peak in the red wing for the near face-on viewing $i$, while the profile is almost constant for the intermediate inclination. We expect that additional chaotic velocity component will affect the profile mostly the core, which is exactly what we get from the models.

In Fig.\,\ref{Contact_ES1_PO} the resulting PO is shown. The degree of polarization is in the same range as it was for the previous case. Again, PO is increasing when viewing from face-on towards edge-on inclinations.

The total flux is largely affected by the additional random motion of the BLR clouds in the line core (Fig.\,\ref{Contact_ES1_TF}). We can clearly observe double-peaked lines for intermediate inclinations ($i = \ang{38}$ and $i = \ang{41}$, Fig.\,\ref{Contact_ES1_TF}, upper panels and bottom left and middle panels). For $\phi = \ang{18},\, \ang{198}$ and $\ang{342}$ (Fig.\,\ref{Contact_ES1_TF}), we observe single-peak profiles, and for intermediate inclinations, line cores are flattened. The highest line widths are for $\phi = \ang{90}$ and $\phi = \ang{270}$.

\subsection{Mixed}
With the two BLRs being mixed and surrounding both black holes, we can observe that the change of $\varphi$ is small with the respect to the continuum level (Figs.\,\ref{f:all} (panel c); \ref{Mixed_ES4_PA}) and it is the highest for nearly face-on inclinations. For intermediate inclinations, the $\varphi$ profiles could be considered as constant with additional noise. This is expected since the largest fraction of flux is coming from the clouds with additional random velocity components that are the close to the black holes.

Figure \ref{Mixed_ES4_PO} shows the resulting PO for a set of viewing inclinations and azimuthal angles. We can see that the broad line profiles are almost flat with very low characteristic features. We obtain the same range for PO as in the previous models.

The total flux is showing seemingly complex profiles (\ref{Mixed_ES4_TF}) with multiple spikes. This is however due to the fact that we are very much limited to the number of BLR clouds when running the simulations. Running the \textsc{stokes} code with more than 5000 individual clouds would be impractical and extremely time consuming. These results are in agreement as the ones obtained by \citet{2005MNRAS.359..846S} i.e.\, we can see that an additional random velocity component besides the Keplerian applied to a large number of BLR clouds, have the tendency to smooth and flatten the resulting spectra. We obtain flat profiles for $\varphi$ and PO, and we can expect a single peaked lines.

\subsection{Spiral} \label{s.spirala}

In Figs.\,\ref{f:all} (panel d); \ref{Spiral_ES6_PA} the results for $\varphi$ for the spiral model are shown. The simulated $\varphi$ is showing double peak profiles whether with minima or maxima occurring around $V \approx \SI{3000}{\kilo\meter\per\second}$ for all $i$ and $\phi$. This velocity is close to the orbital velocity of each binary component for which $V \approx \SI{2800}{\kilo\meter\per\second}$. This result is due to most of the emitted flux that is originating from the inner parts of the spiral arms closer to the black holes, and due to the velocity of the rigid body scaling with the distance. The intensity of the peaks is inclination dependent and is decreasing when the system is viewed from face-on towards edge-on inclinations.

In Fig.\,\ref{Spiral_ES6_PO} the results for the simulated PO are shown. We can see that PO is having similar profiles as $\varphi$ -- visible peaks in the line wings and minimum in the line core (Fig.\,\ref{Spiral_ES6_PO}, left upper and middle panels; right bottom and middle panels) that is characteristic for a single black hole scenario, or the opposite profiles with maximum PO in the line core and minimum in the wings.

The results for TF are shown in Fig.\,\ref{Spiral_ES6_TF}. We can see various line profiles for different $\phi$. For intermediate inclinations, we observe double-peaked line profiles. For near face-on viewing angles and some $\phi$, profiles with strong single peak (Fig.\,\ref{Spiral_ES6_TF}, bottom right panel), or two peaks very close to each other (Fig.\,\ref{Spiral_ES6_TF}, middle left and right panels) are observed.

\section{Discussion} \label{S.discussion}

\subsection{Overview of our results}
The presence of another BLR (as in the case of our model) has a unique signature on the simulated $\varphi$-profiles for all the models we tested. A double peaked feature can be observed, and the $\varphi$-profile is varying drastically depending on the observed orbital phase of the system and it is different than in case of a single SMBH. This is always the case for PO and TF, which often show complex profiles. However, in some cases, when viewed from certain azimuthal viewing angles, the simulated PO and TF profiles are very similar for the case with a single SMBH in the center. AGN monitoring is therefore required for distinguishing between these two cases. We have seen that additional random motion tends to smooth the profiles of TF in the line core, while diluting $\varphi$-profiles. The total flux is also largely dependent on the observed phase of the binary system. Lines show complex varying profiles, and long-term monitoring spectroscopy, combined with spectropolarimetry could prove very useful in the search for SMBBH candidates. In order to see the variability in the line, we are limited only to close subparsec SMBBHs for which the half-period of revolution is of the order up to few tens of years. Less massive SMBBHs or the ones with greater orbital distance would yield orbital periods of the order of few centuries, that the line profile change would be impractical to observe.

In \citet{2018A&A...614A.120S}, we simulated equatorial scattering with additional complex (inflows/outflows) motion in the BLR with for a single SMBH in the center. The $\varphi$-profiles are showing point (central) symmetry for all treated cases (e.g.\,a prominent minimum followed by a maximum of the same amplitude), while the TF remains axisymmetric with the respect to the line center. \citet{2005MNRAS.359..846S} have included inflows and high-velocity rotation in the scattering region and it yielded complex, but again point symmetric $\varphi$-profiles. Depending on the model, our simulations involving SMBBHs as a result have axisymmetric $\varphi$-profiles. This behavior of polarization angle may prove crucial as a distinct feature in the search for SMBBHs.

\subsection{AGNs with double-peaked emission line profiles}
Broad emission line profiles and line variability can be explained by a wide variety of different kinematic models that would yield similar results. Naturally, AGNs with variable double-peaked lines make good targets for spectropolarimetric observations and long-term monitoring campaigns. We discuss our results with observations of three well known double-peaked AGNs: NGC 1097, 3C 390.3 and Arp 102B.

Spectral optical monitoring of Arp 102B over the period from 1987 to 2010 shows no significant change in the broad double-peaked H$\alpha$ and H$\beta$ profiles \citep{2013A&A...559A..10S,2014A&A...572A..66P}. The H$\beta$ line is broader than H$\alpha$ during the monitored period and both can be well reproduced by disk model. However, spectropolarimetric observations are partially inconsistent with the disc model \citep{1998MNRAS.296..721C,2000MNRAS.319..685C}. The H$\alpha$ polarization angle has almost the same value as the angle of the jet direction, which is in good agreement with equatorial scattering. The observed single-peak profile of the polarized line with respect to the unpolarized suggest that the BLR clouds might be undergoing biconical outflows \citep{1996ApJ...456L..25A,1998MNRAS.296..721C,2000MNRAS.319..685C}. The $\varphi$-profile is flat without any distinctive feature.

The active galaxy 3C 390.3 is a well known source with remarkably strong variability in the X-, UV and optical regime \citep[see][and the references therein]{2015MNRAS.448.2879A}. The unpolarized and polarized flux are quite different. The unpolarized H$\alpha$ has a double-peaked profile with blue peak being more prominent. The polarized H$\alpha$ is single-peaked shifted to blue for \SI{1200}{\kilo\meter\per\second} with the respect to narrow component and is strongly depolarized in the center \citep{2015MNRAS.448.2879A}. A model with biconical outflows \citep{2000MNRAS.319..685C} for this object is not in agreement with the optical monitoring of the BLR. The CCF analysis by \citet{2015MNRAS.448.2879A} for H$\alpha$ and H$\beta$ shows no significant delay in the variation between the blue and the red line wing relative to each other or with the respect to the line core. This is in favour of a model with the BLR originating from an accretion disk with dominant Keplerian motion. A two-component BLR model with disc and an outflowing region can well explain spectropolarimetric observations. In this model, an outflowing region is located above the disk and it can depolarize the radiation emitted from the disk.

Optical monitoring of NGC 1097 between 1991 and 1996 have shown a peculiar evolution of the H$\alpha$ line profile. The broad H$\alpha$ double peak showed a red-peak dominance \citep{1993ApJ...410L..11S}, followed by a nearly symmetrical profile \citep{1995ApJ...443..617S} and up to a blue-peak dominant profile \citep{1997ApJ...489...87S}. A model of a precessing elliptical ring around the SMBH \citep{1995ApJ...438..610E} was used to explain observed line profiles and to fit the data. In this model it was proposed that the origin of the elliptical disk is due to the tidal disruption of a star by a SMBH or it could be due to the existence of a SMBBH. In both cases, broad line variability that could be observed is of the order of few years when the total mass is smaller than \SI{d6}{\solarmass}. For SMBH with mass of the order of \SI{d8}{\solarmass}, which is the case for NGC 1097\footnote{\\$\ind{M}{bh}$ = \SI[separate-uncertainty = true]{1.2(2)e8}{\solarmass} \citep{2006ApJ...642..711L};\\ $\ind{M}{bh}$ = \SI[separate-uncertainty = true]{1.40(32)e8}{\solarmass} \citep{2015ApJ...806...39O}}, the precession period is of the order of a few centuries and could not be observed. However, in the scenario we studied, where each component is having a separate accretion disk surrounded by the BLR, the variability of the order of few years could be observed if the binary system is close enough. Line variability would show systematic periodicity and it is attributed only to the viewed orbital phase of the system. In order to fit the observational data with our model, a large grid of models needs to be conducted. Besides the main parameters of the model such as total mass, orbital distance and mass ratio, the parameter space would also include luminosities and BLR sizes of each components along with the parameters describing the scattering region as well as the optical depth. This is well beyond the scope of the present work and limits our investigation based on a simple model.

When viewed in polarized light, NGC 1097 shows a weak continuum polarization ($p = 0.26 \pm \SI{0.02}{\percent} $) in optical domain over \SI{5100}{}--\SI{6100}{\angstrom} \citep{1999ApJ...525..673B}. The H$\alpha$ line is also showing weak polarization and no characteristic feature for a single or binary BH could be detected in the PO and PF profiles. New high quality spectropolarimetric observations are thus required in order to confirm our results.

AGNs with unpolarized double-peaked profiles with varying red and blue peak with respect to each other are probably the best candidates in the search for SMBBHs. Although a single SMBH in the center of AGNs is the most probable case, SMBBH in the central engine should have their distinctive signiture in the polarized spectra due to the polarization sensitivity on geometry and kinematics.

\section{Conclusions} \label{S.conclusions}
We investigated the polarization signatures of SMBBHs in AGNs using a set of simple yet representative models. We assumed equatorial scattering as a main mechanism for optical polarization and we used the Monte Carlo code \textsc{stokes} for solving 3D polarized radiative transfer with kinematics. We used simple geometry for polarization modeling of SMBBHs in AGNs and we treated four different cases with different geometry of the BLRs: \textbf{distant}, \textbf{contact}, \textbf{mixed} and \textbf{spiral}. We outline the characteristic features of $\varphi$, PO and TF that are in common for all the models we studied. Polarization position angle $\varphi$ is showing double-peaked or even more complex profiles most of the time. The PO shows double-peak profiles with minimum in the line core, which is common for the single SMBH scenario, but there are opposite profiles with minima in the line wings and maximal PO in the line core which may be an indicator of a SMBBHs. The TF shows most of the time double-, or multi-peaked profiles which are often associated with the disk profiles. The combined results of all of our simulations involving SMBBHs leads to the following two conclusions:

\begin{itemize}
 \item The degree of polarization and total flux, along with the unique profiles characteristic for SMBBHs also show profiles that are common for single SMBHs and alone may prove inconclusive for disentangling the central engine of AGNs.
 \item On the other hand, the polarization position angle $\varphi$ shows quite unique profiles than the ones observed for single SMBH scenario, and it's inspection could be used as a first step for finding the SMBBH candidates.
\end{itemize}

We demonstrated that when a SMBBH is situated in the center of Type-1 AGNs, spectropolarimetry could be a powerful tool for searching the SMBBH candidates amongst them. In this paper we assumed that the accretion disks of the two black holes are coplanar and that they are coplanar with the torus, i.e.\,scattering region.  Our assumption of coplanarity is very well supported by previous results from high-resolution hydrodynamical simulations. However, for a general picture of how significant is the orientation between the disks in the short-lived phase with misaligned disks, a more detailed analysis with the expanded model space grid is required. With the results obtained so far, in this case, we expect to have highly asymmetric profiles of the total flux and the degree of polarization, while for the polarization position angle, we expect to have lower amplitude and more flat profiles. We intend to explore the cases of non-coplanarity in a follow-up paper that will investigate the whole (and large) phase space of free parameters.

\begin{acknowledgements}
      This work was supported by the Ministry of Education and Science (Republic of Serbia) through the project Astrophysical Spectroscopy of Extragalactic Objects (176001), the French PNHE and the grant ANR-11-JS56-013-01 “POLIOPTIX”. F.\,M.\,is grateful to the Centre national d'études spatiales (CNES) and its post-doctoral grant "Probing the geometry and physics of active galactic nuclei with ultraviolet and X-ray polarized radiative transfer". D.\,Savi\'c thanks the French Government and the French Embassy in Serbia for supporting his research without which this work would not be possible.
\end{acknowledgements}

\begin{appendix}
\section{Detailed results of modeling}
\begin{figure*}
   \centering
   \includegraphics[width=\hsize]{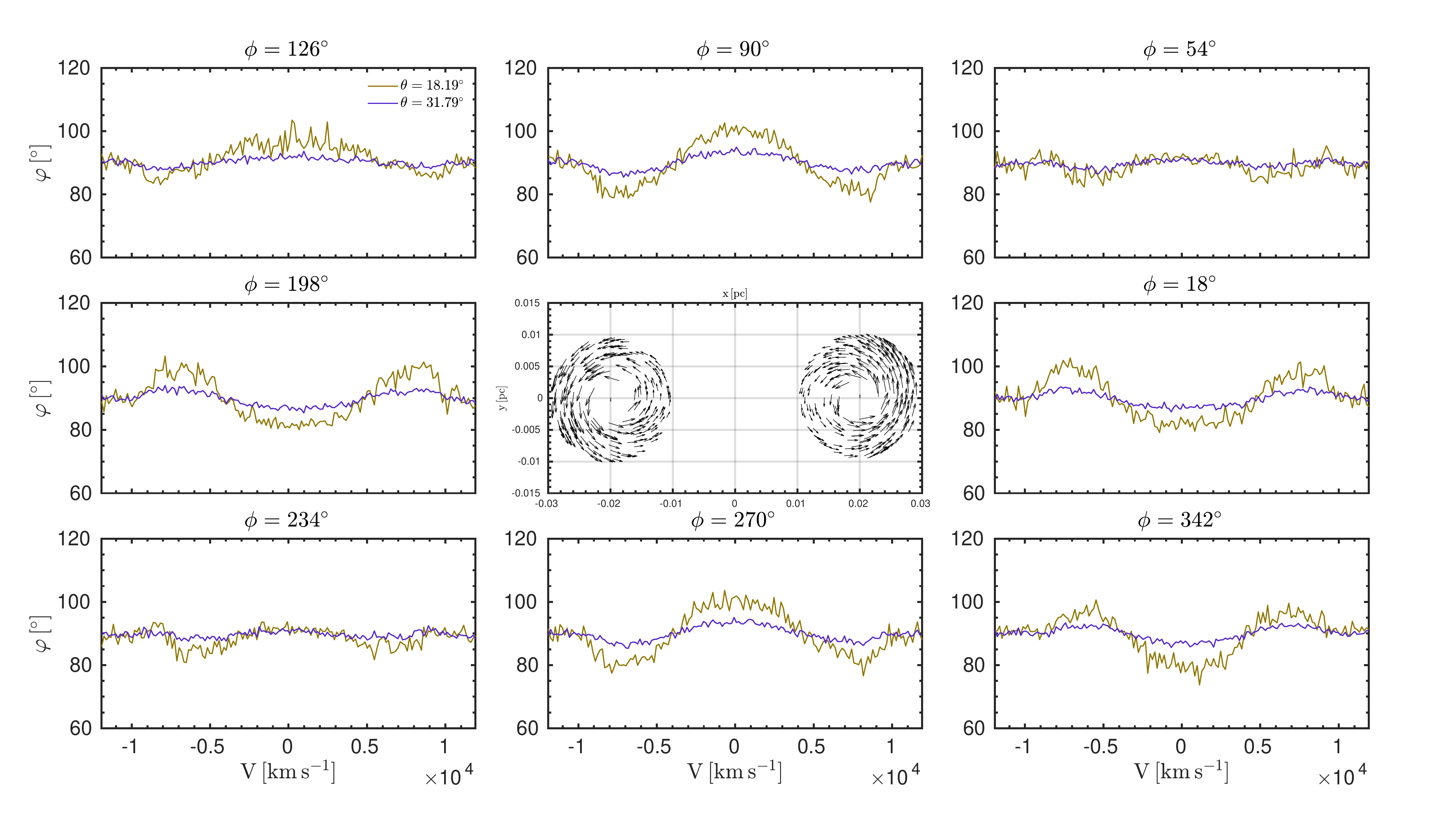}
      \caption{Simulated profiles of $\varphi$ across the line profile for two viewing inclinations $i$ when observed from different azimuthal angles $\phi$. Geometry and kinematics of the model is in the center for clarity.}
         \label{Distant_ES12_PA}
\end{figure*}

\begin{figure*}
   \centering
   \includegraphics[width=\hsize]{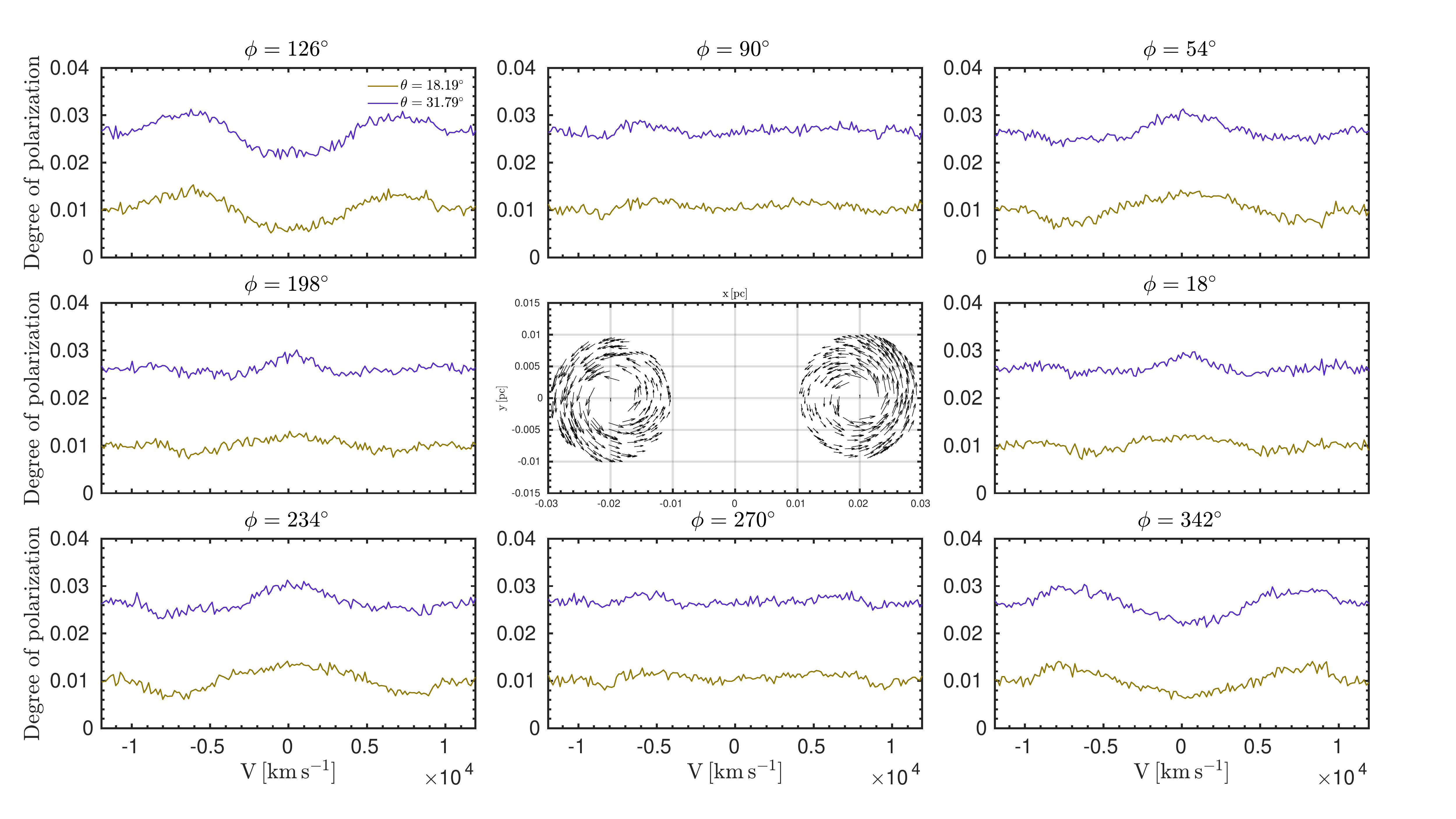}
      \caption{Same as figure \ref{Distant_ES12_PA}, but for PO.}
         \label{Distant_ES12_PO}
\end{figure*}

\begin{figure*}
   \centering
   \includegraphics[width=\hsize]{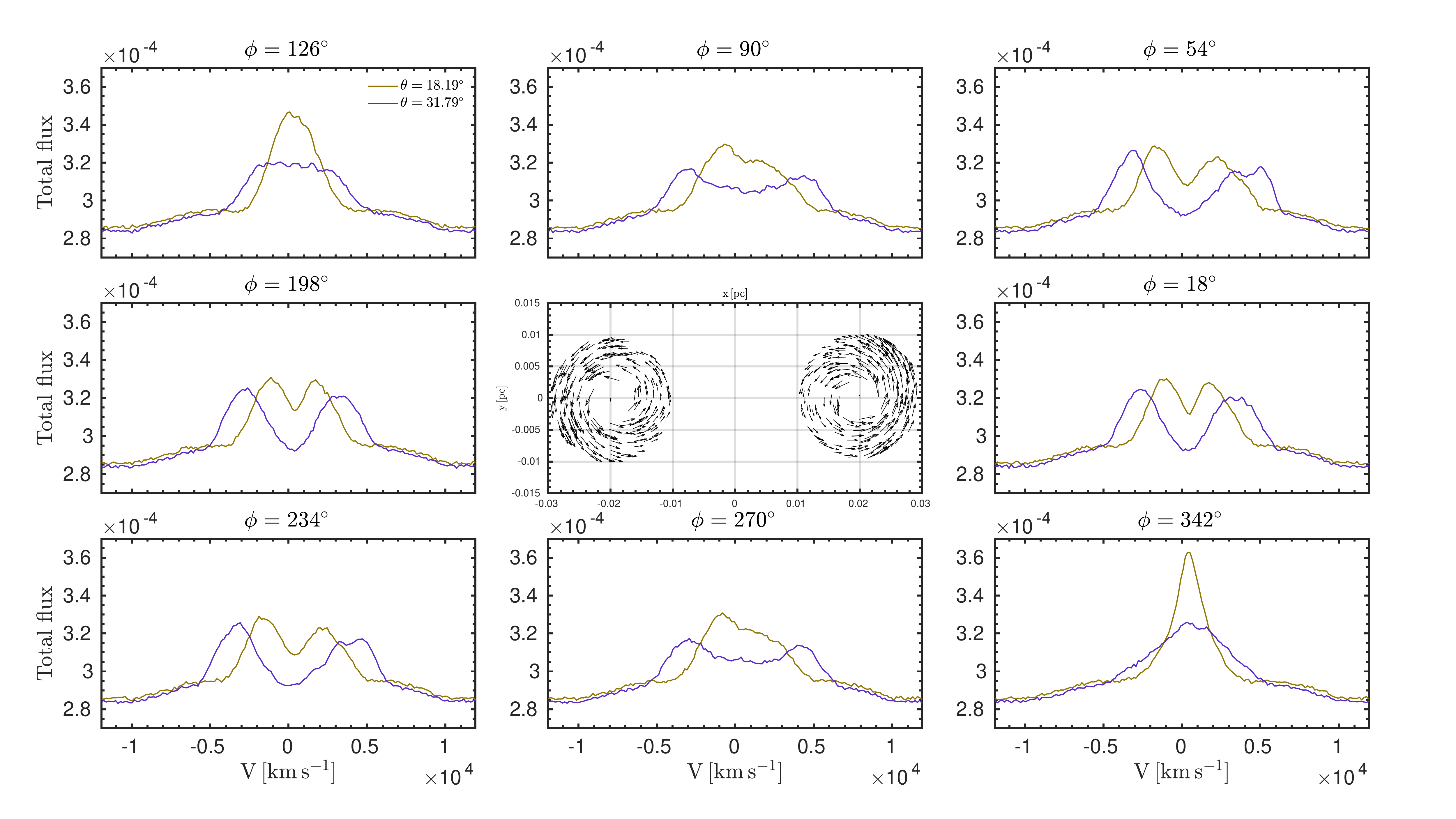}
      \caption{Same as figure \ref{Distant_ES12_PA}, but for TF.}
         \label{Distant_ES12_TF}
\end{figure*}

\begin{figure*}
   \centering
   \includegraphics[width=\hsize]{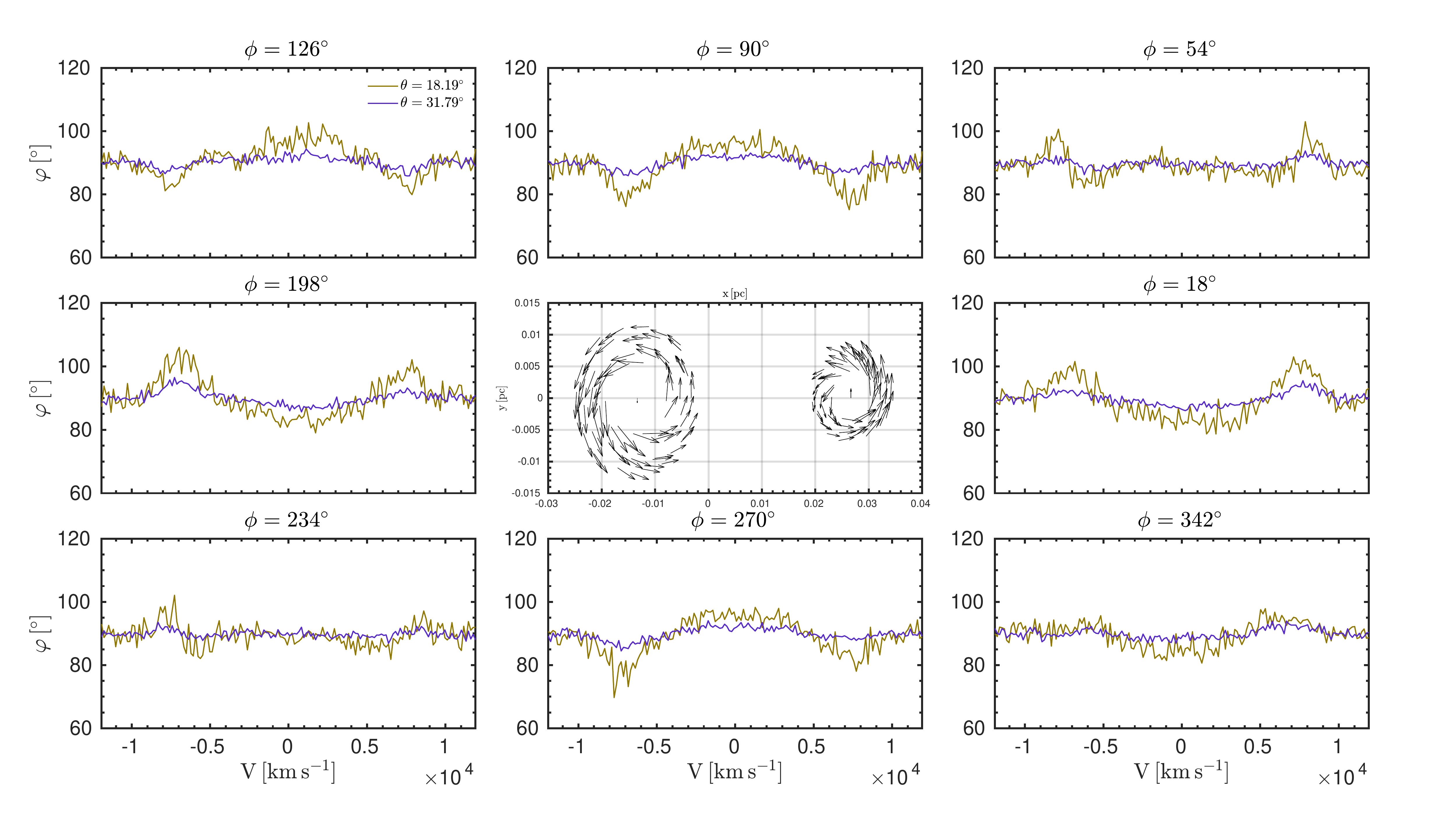}
      \caption{Same as figure \ref{Distant_ES12_PA}, but for $q = 0.5$.}
         \label{Distant_ES12_q05_PA}
\end{figure*}

\begin{figure*}
   \centering
   \includegraphics[width=\hsize]{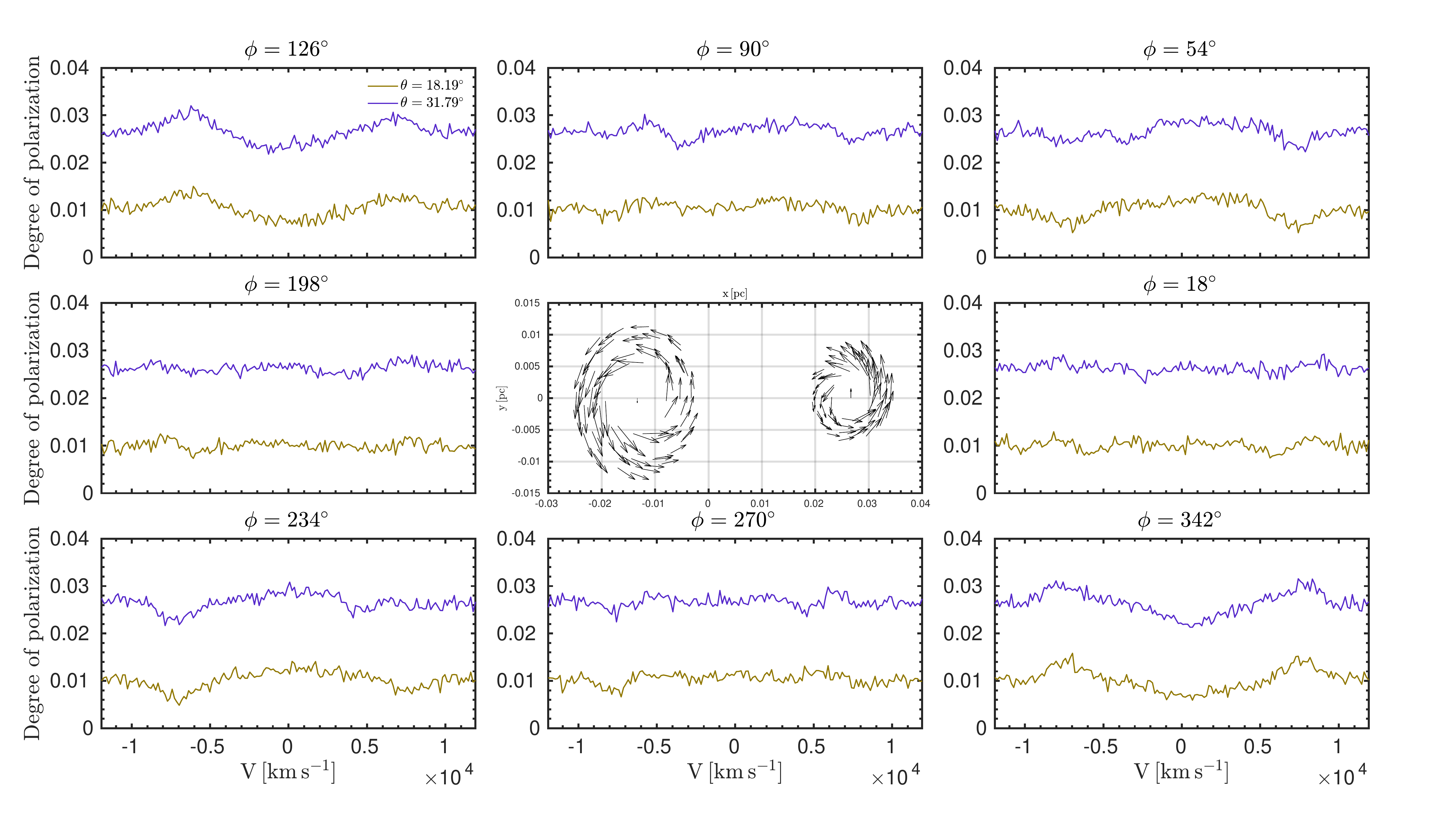}
      \caption{Same as figure \ref{Distant_ES12_PO}, but for $q = 0.5$.}
         \label{Distant_ES12_q05_PO}
\end{figure*}

\begin{figure*}
   \centering
   \includegraphics[width=\hsize]{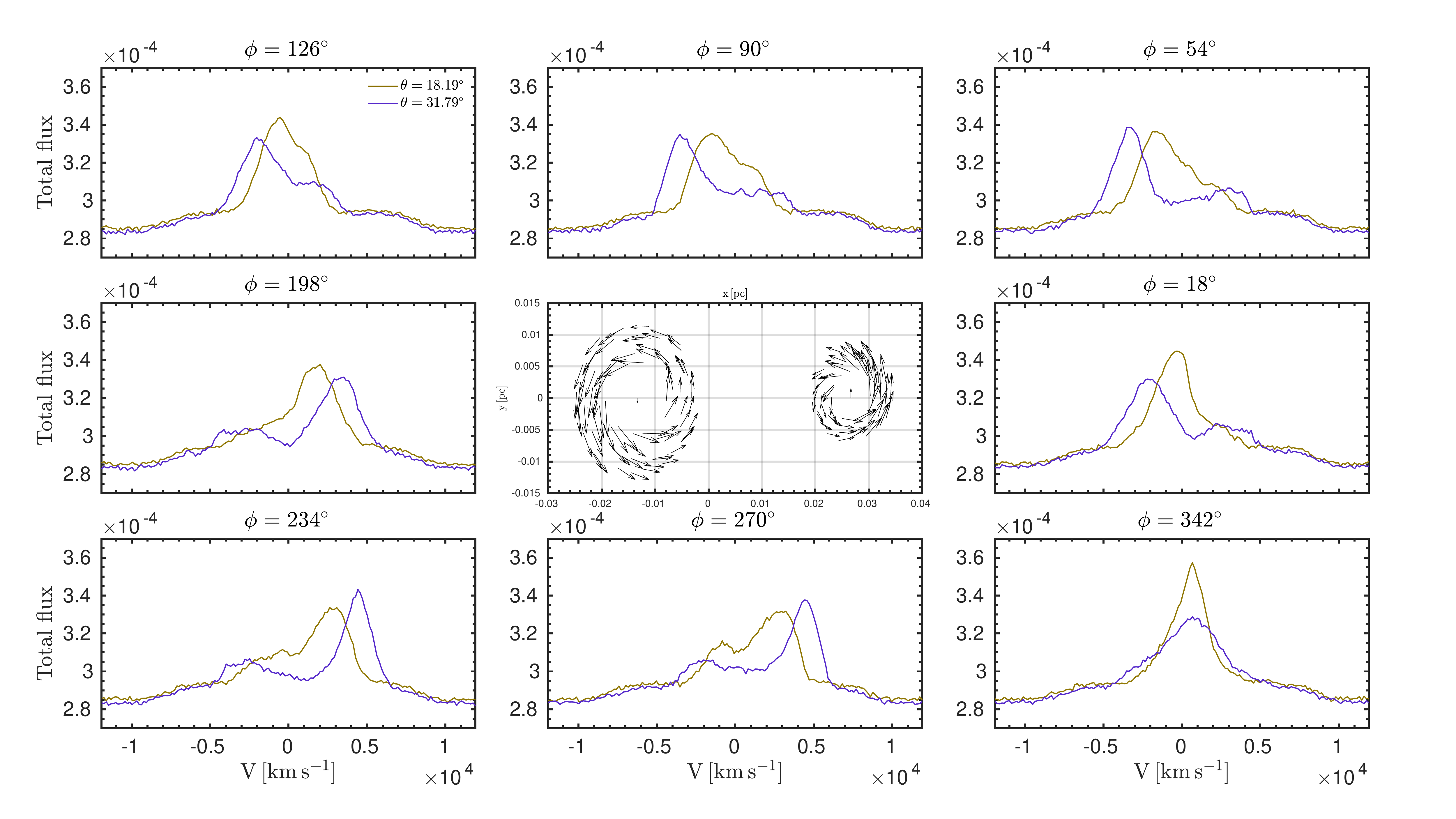}
      \caption{Same as figure \ref{Distant_ES12_TF}, but for $q = 0.5$.}
         \label{Distant_ES12_q05_TF}
\end{figure*}

\begin{figure*}
   \centering
   \includegraphics[width=\hsize]{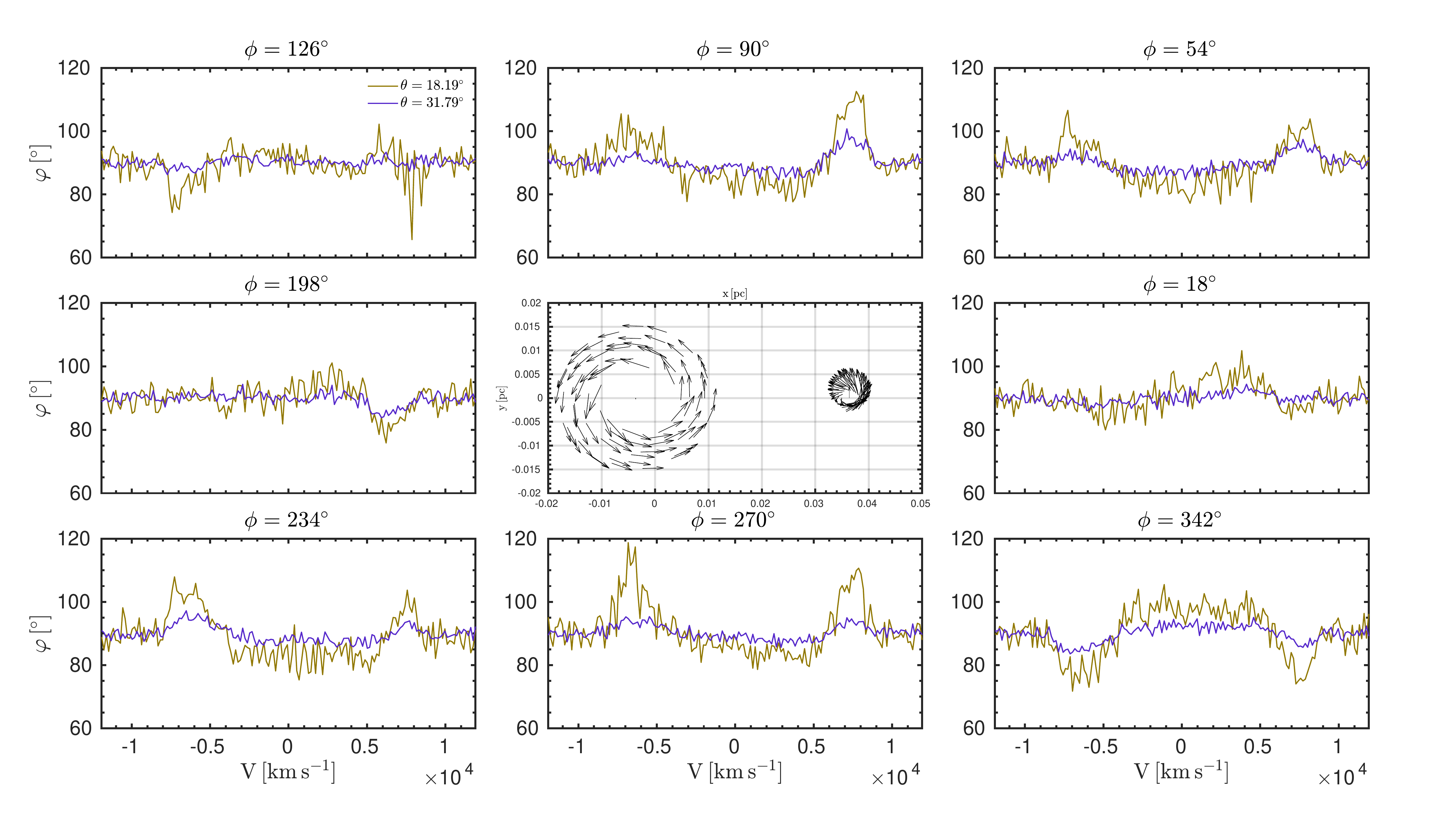}
      \caption{Same as figure \ref{Distant_ES12_PA}, but for $q = 0.1$.}
         \label{Distant_ES12_q01_PA}
\end{figure*}

\begin{figure*}
   \centering
   \includegraphics[width=\hsize]{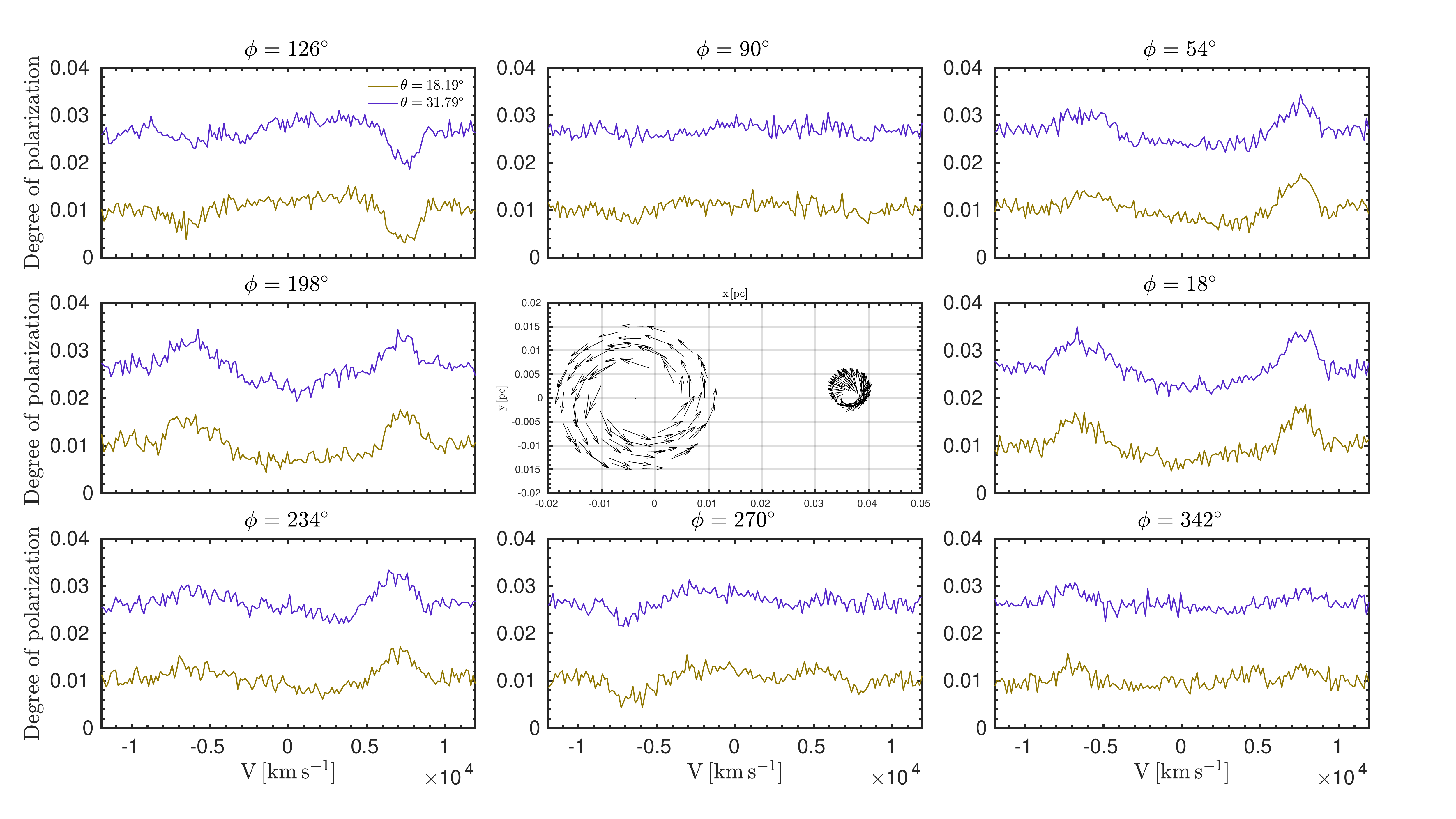}
      \caption{Same as figure \ref{Distant_ES12_PO}, but for $q = 0.1$.}
         \label{Distant_ES12_q01_PO}
\end{figure*}

\begin{figure*}
   \centering
   \includegraphics[width=\hsize]{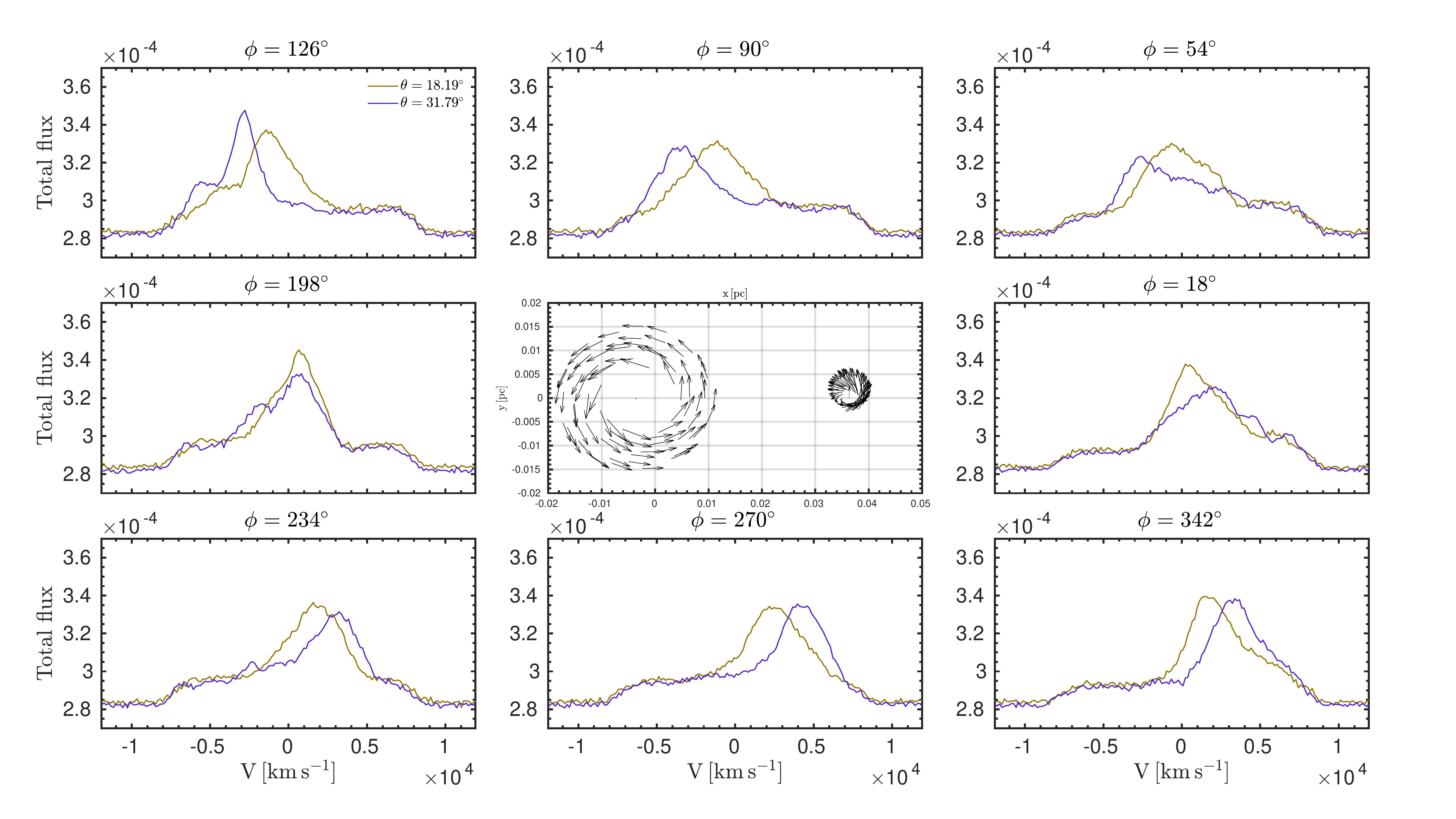}
      \caption{Same as figure \ref{Distant_ES12_TF}, but for $q = 0.1$.}
         \label{Distant_ES12_q01_TF}
\end{figure*}

\begin{figure*}
   \centering
   \includegraphics[width=\hsize]{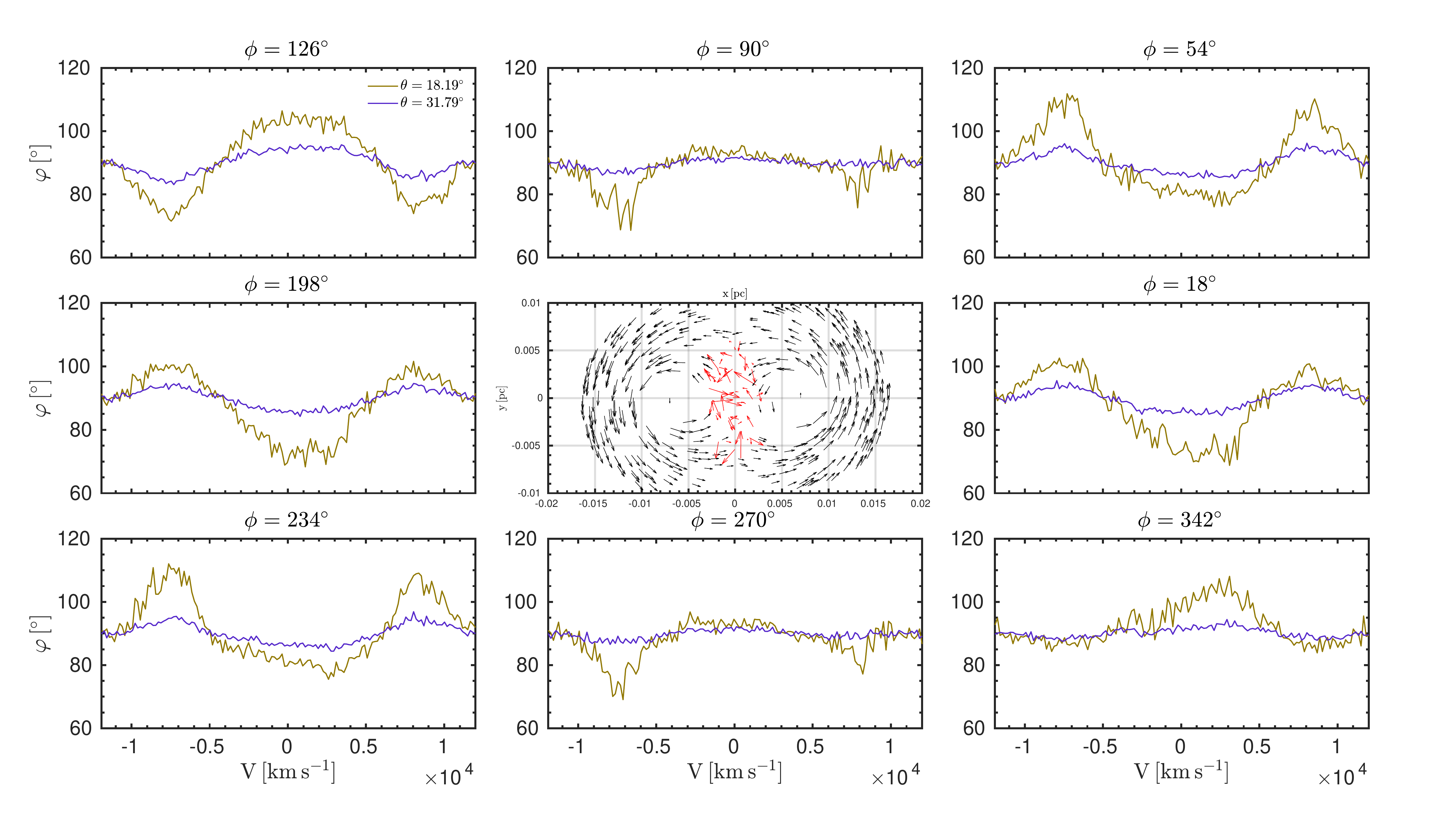}
      \caption{Same as figure \ref{Distant_ES12_PA}, but for contact model.}
         \label{Contact_ES1_PA}
\end{figure*}

\begin{figure*}
   \centering
   \includegraphics[width=\hsize]{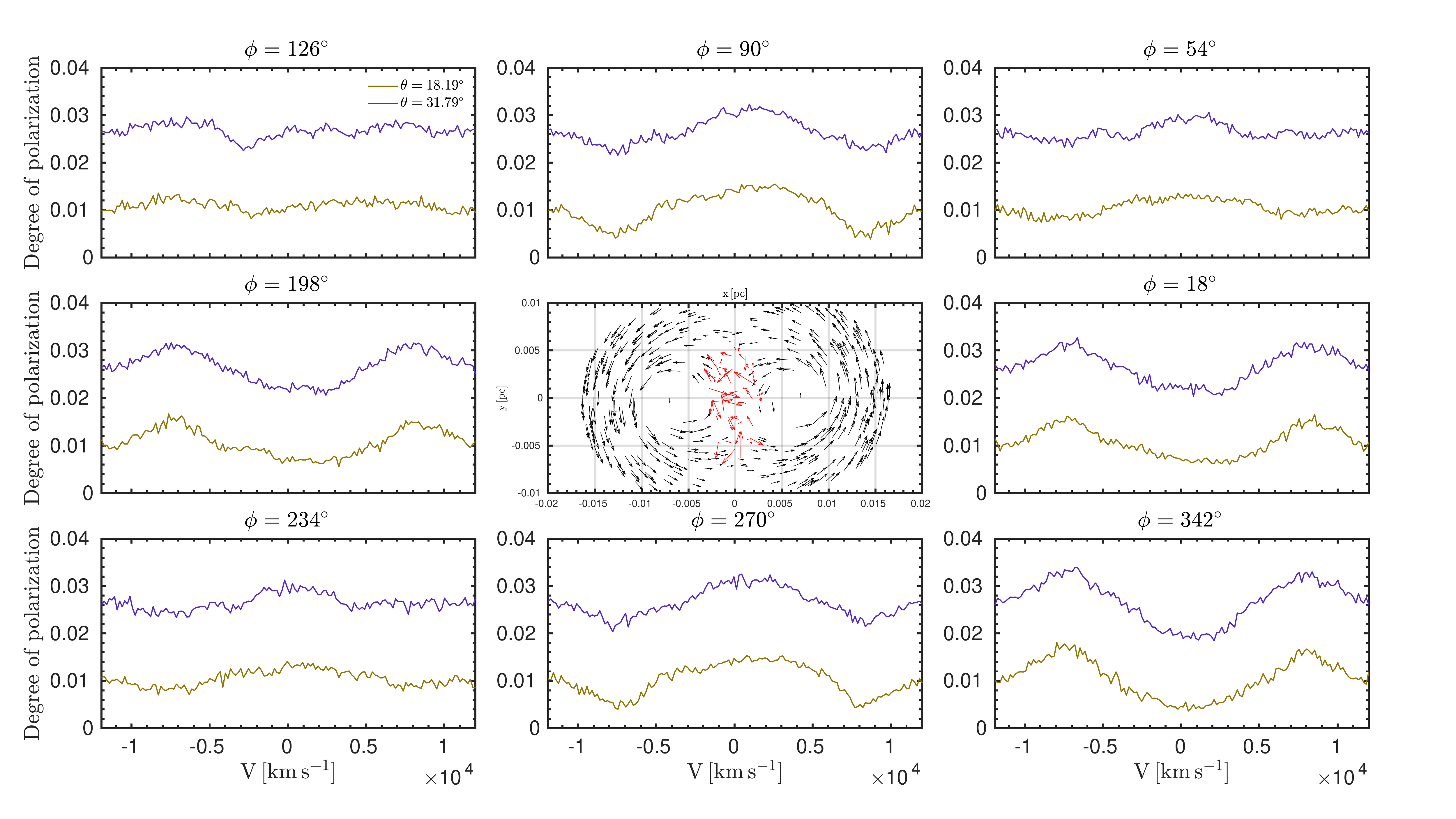}
      \caption{Same as figure \ref{Contact_ES1_PA}, but for PO.}
         \label{Contact_ES1_PO}
\end{figure*}

\begin{figure*}
   \centering
   \includegraphics[width=\hsize]{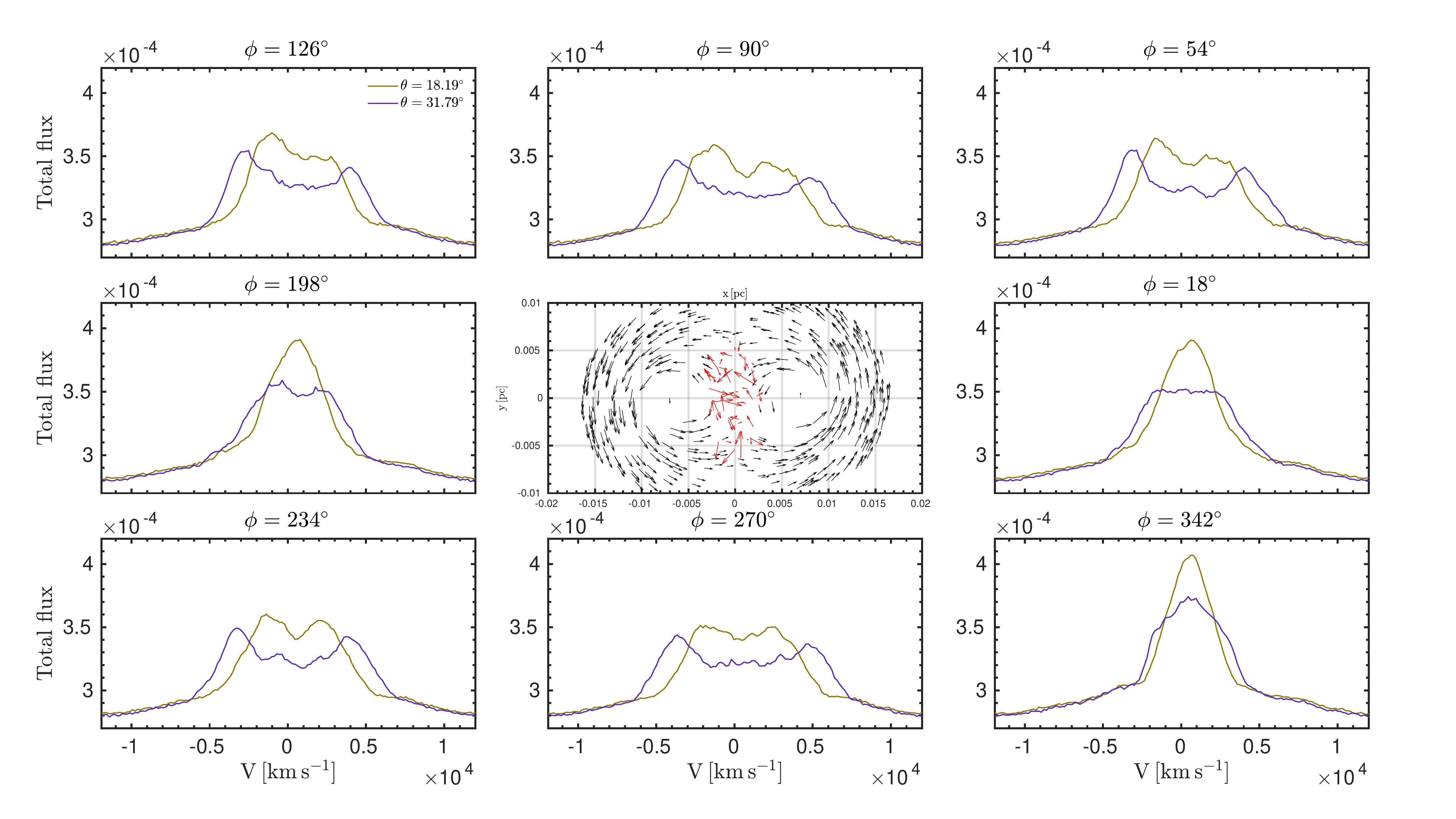}
      \caption{Same as figure \ref{Contact_ES1_PA}, but for TF.}
         \label{Contact_ES1_TF}
\end{figure*}

\begin{figure*}
   \centering
   \includegraphics[width=\hsize]{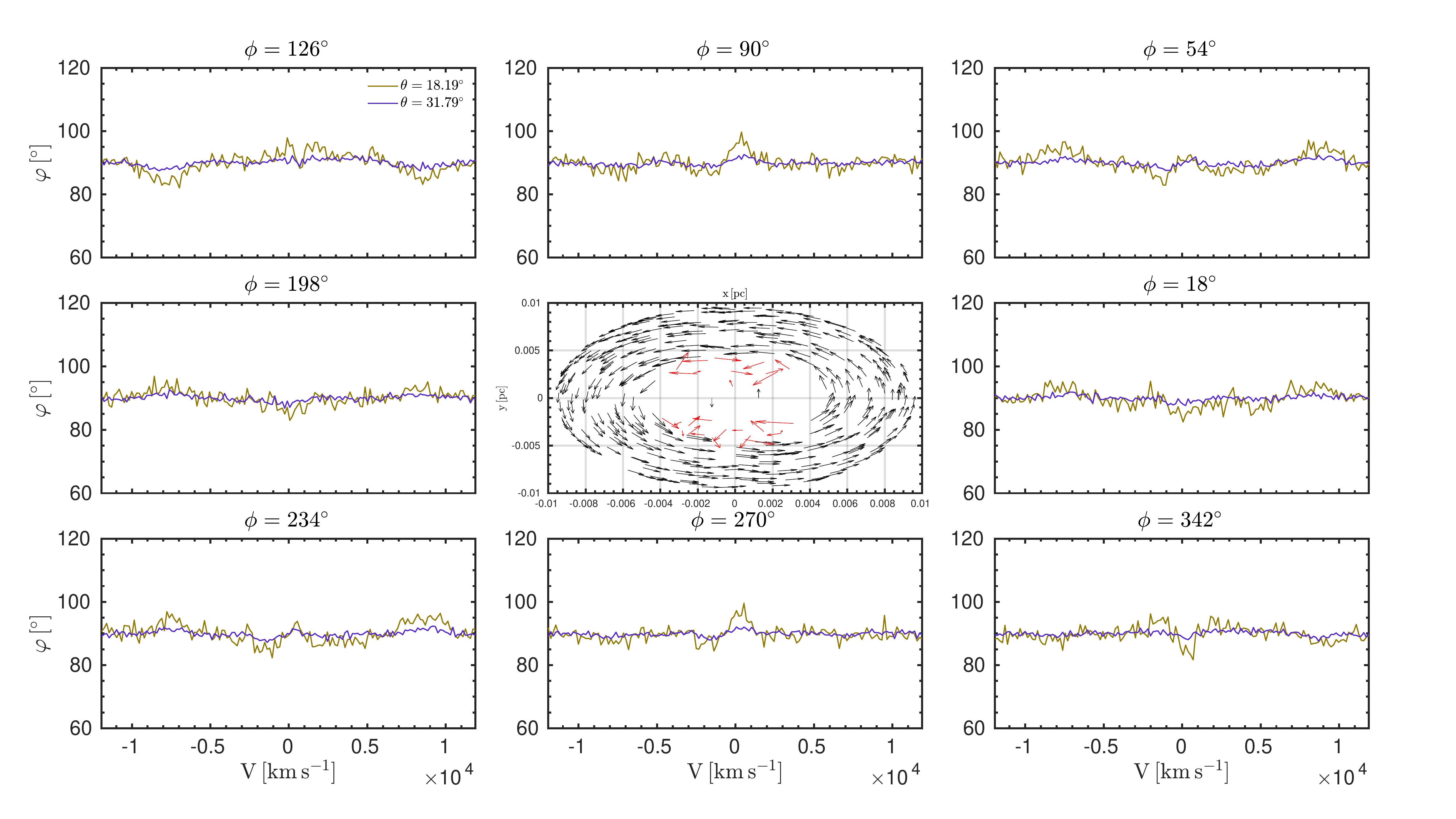}
      \caption{Same as figure \ref{Distant_ES12_PA}, but for mixed model.}
         \label{Mixed_ES4_PA}
\end{figure*}

\begin{figure*}
   \centering
   \includegraphics[width=\hsize]{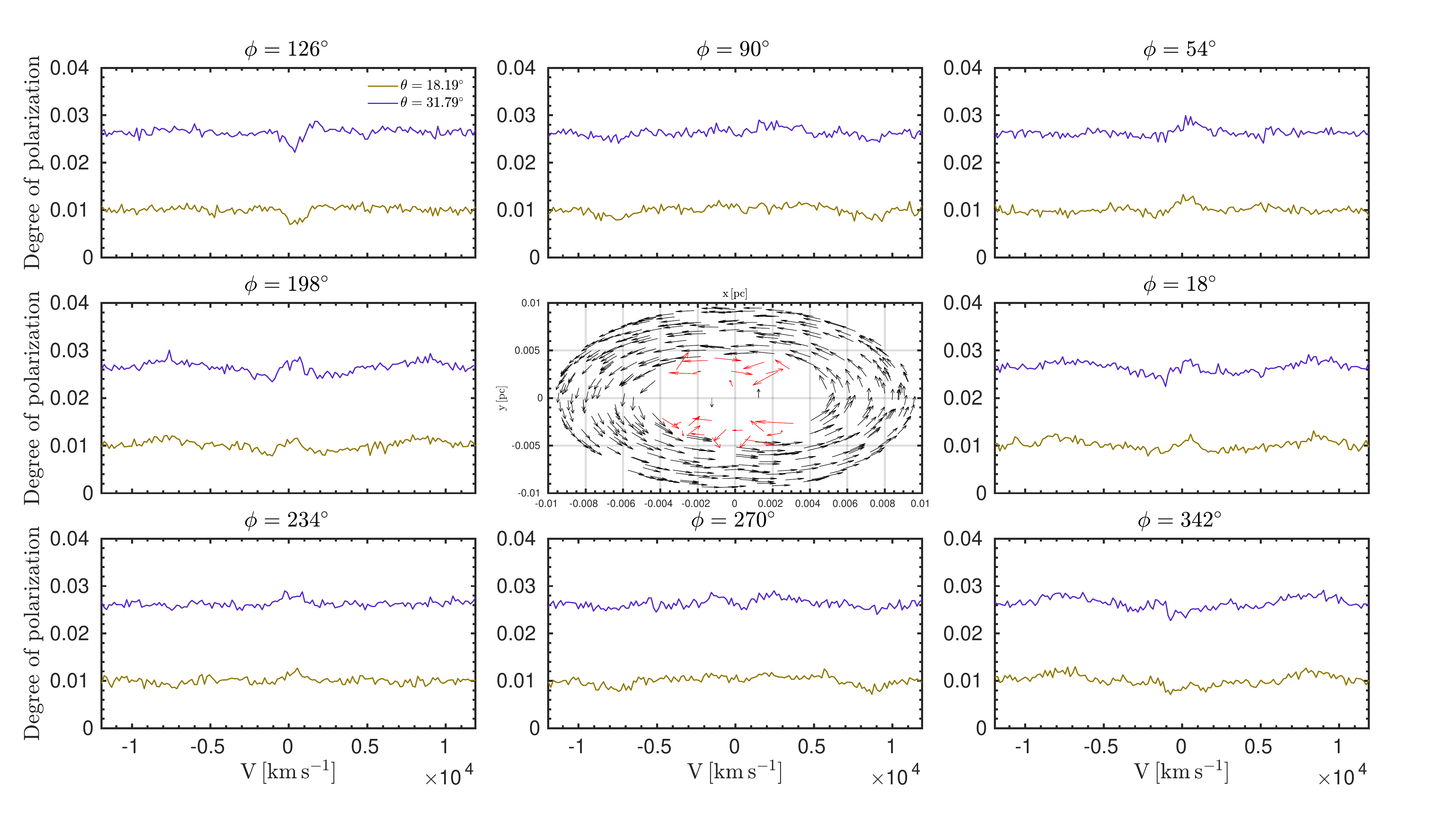}
      \caption{Same as figure \ref{Mixed_ES4_PA}, but for PO.}
         \label{Mixed_ES4_PO}
\end{figure*}

\begin{figure*}
   \centering
   \includegraphics[width=\hsize]{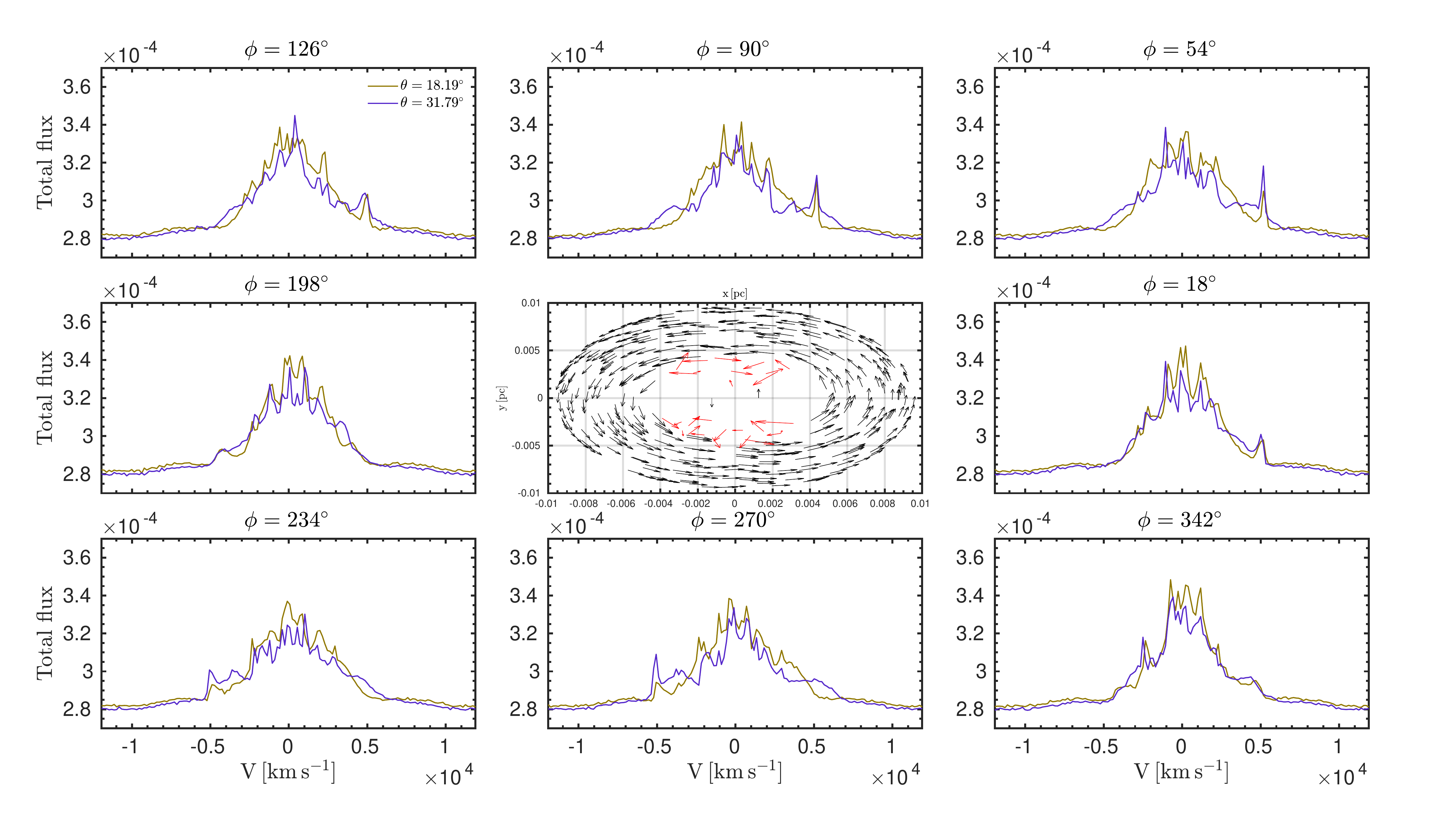}
      \caption{Same as figure \ref{Mixed_ES4_PA}, but for TF.}
         \label{Mixed_ES4_TF}
\end{figure*}

\begin{figure*}
   \centering
   \includegraphics[width=\hsize]{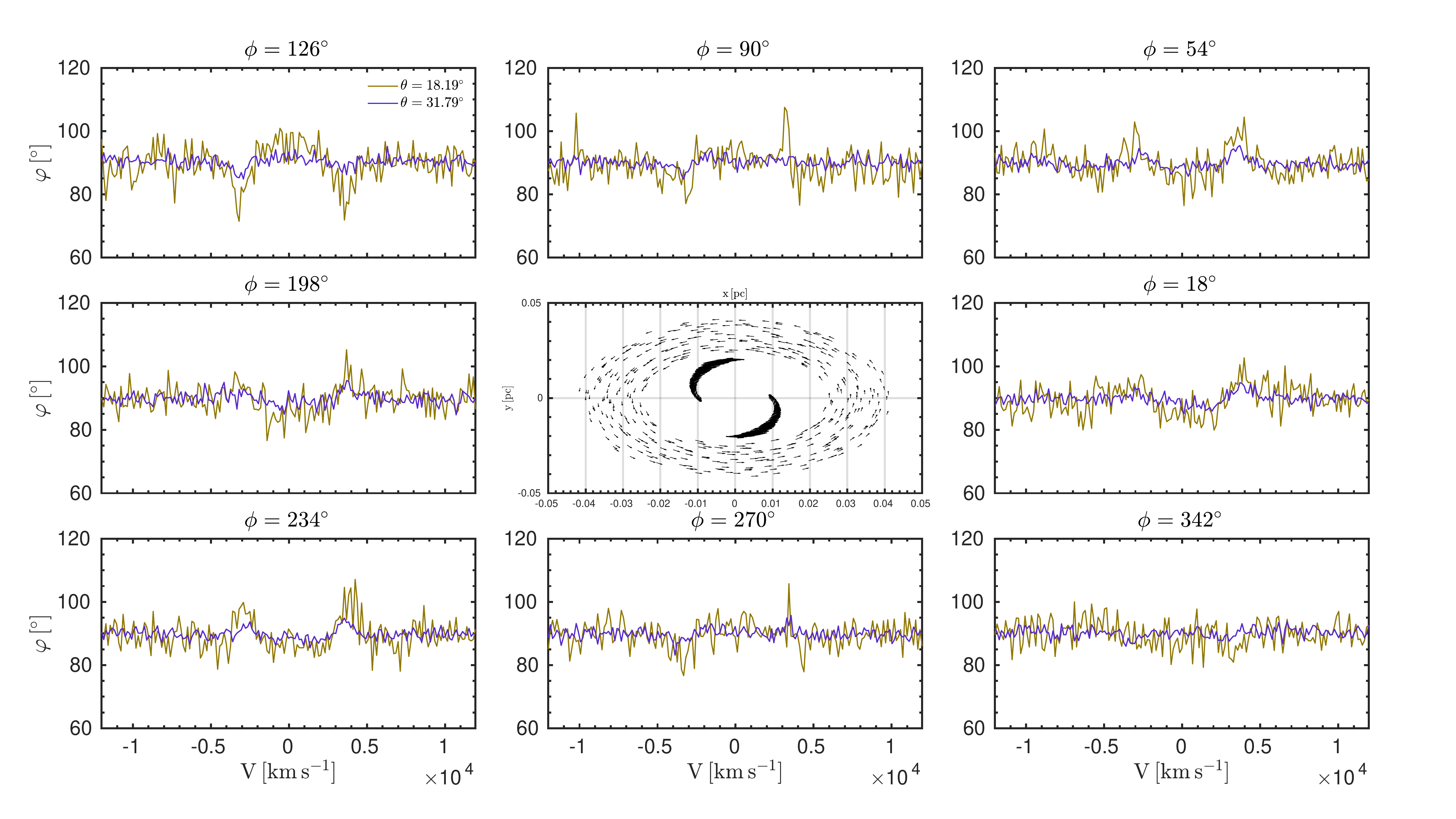}
      \caption{Profiles of $\varphi$ across the line profile.}
         \label{Spiral_ES6_PA}
\end{figure*}

\begin{figure*}
   \centering
   \includegraphics[width=\hsize]{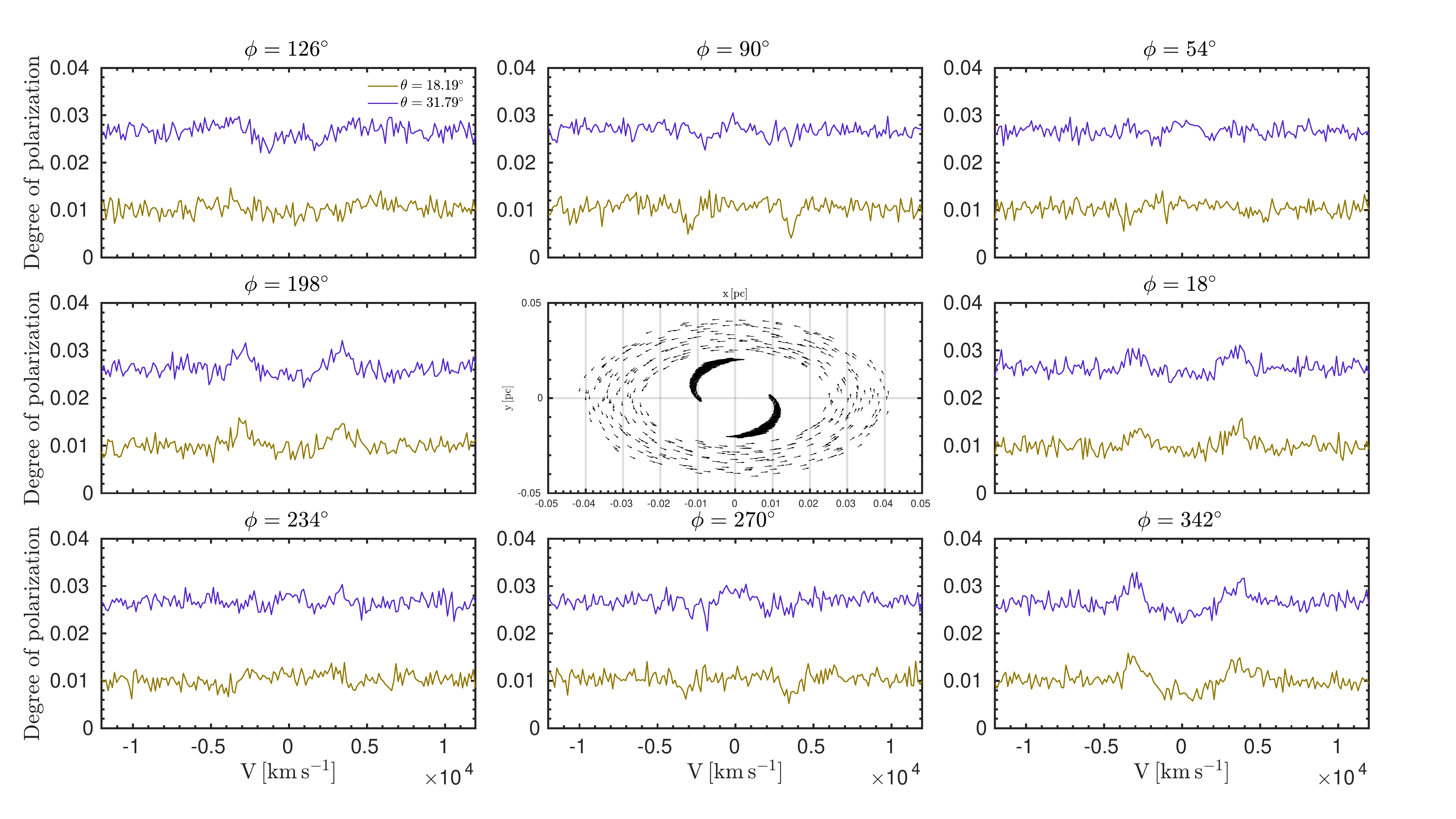}
      \caption{Profiles of PO across the line profile.}
         \label{Spiral_ES6_PO}
\end{figure*}

\begin{figure*}
   \centering
   \includegraphics[width=\hsize]{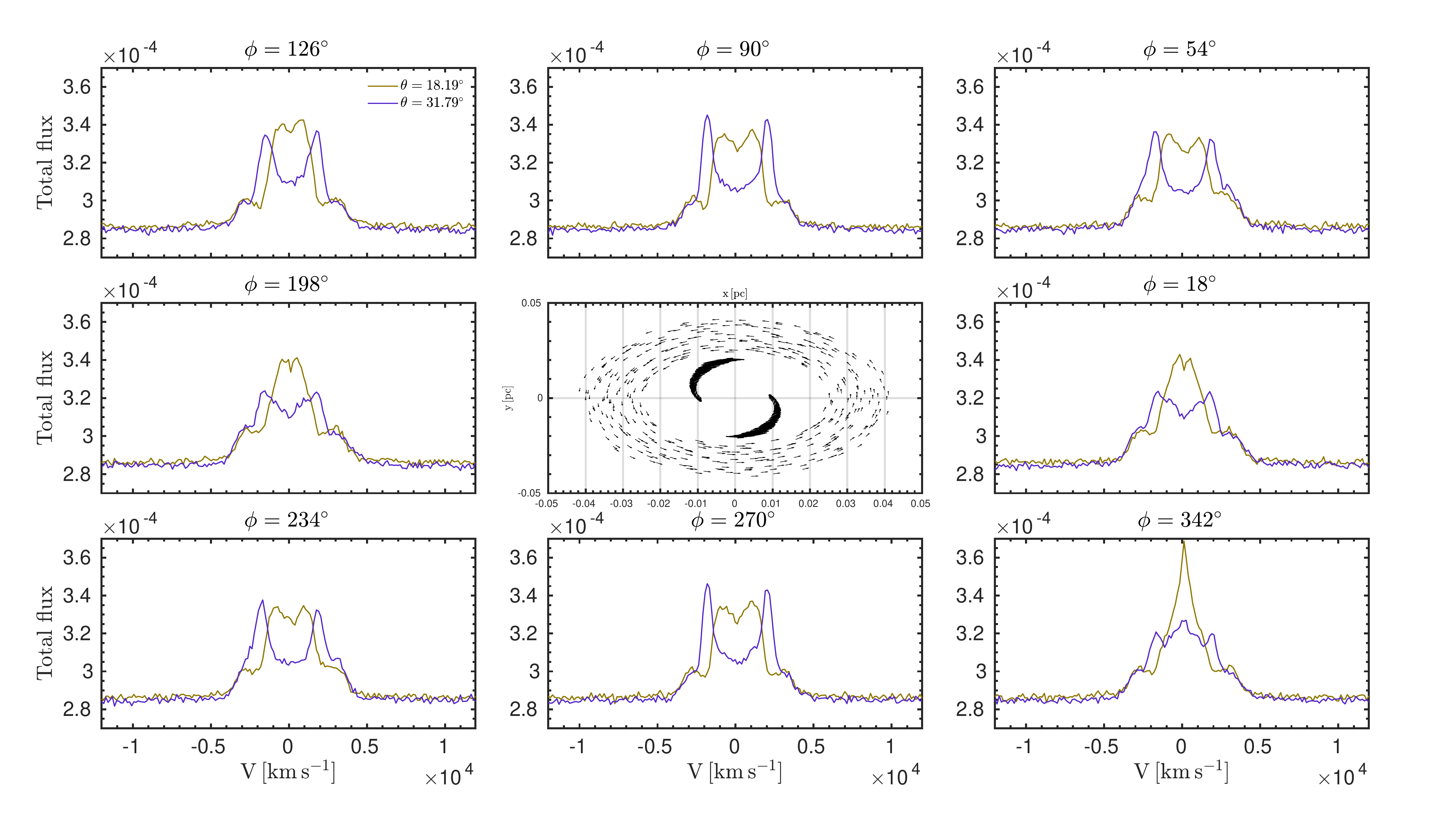}
      \caption{Profiles of TF across the line profile.}
         \label{Spiral_ES6_TF}
\end{figure*}

Simulations for all models are presented in the figures from \ref{Distant_ES12_PA} to \ref{Spiral_ES6_TF}. The simulated profiles for $\varphi$, PO and TF are given for two viewing inclinations: $i \approx \ang{18}$ and $\ang{32}$. Azimuthal viewing angles takes eight values: $\phi = \ang{18},\, \ang{54},\, \ang{90},\, \ang{126},\, \ang{198},\, \ang{224},\, \ang{270}$ and $\ang{342}$. The results are given as a function of velocity defined as $V = c(\lambda - \lambda_0)/\lambda_0$, where $\lambda$ is wavelength  and $\lambda_0$ is the central wavelength of a given spectral line. The broad line region for each model is shown in the center of every image. Arrows represent the velocity field of the BLR. For each model, we outline the main features for completeness.

\textbf{Distant}: This case is shown in Figs.\,\ref{Distant_ES12_PA}-\ref{Distant_ES12_TF} for mass ratio $q = 1$. In Fig.\,\ref{Distant_ES12_PA} the polarisation angle $\varphi$ is shown. We can observe a double-peaked profiles of $\varphi$ that drastically vary depending on the orbital phase of the system. For $\phi = \ang{18}$ and $\ang{198}$, $\varphi$ reaches maximum values in the line wings and minimum in the core, while for $\phi = \ang{90}$ and $\phi = \ang{270}$, it is the opposite way around. The PO is shown in Figs.\,\ref{Distant_ES12_PO} shows similar profiles as $\varphi$, but they are not correlated. Profiles with minimum in the core and maxima in the wings, which is common for the single SMBH scenario can be seen for $\phi = \ang{126}$ and $\ang{342}$. The opposite profiles are for $\phi = \ang{18},\,\ang{54},\,\ang{198}$ and $\ang{234}$. The TF is shown in Fig.\,\ref{Distant_ES12_TF}. Double-peaked profiles can be seen for $\phi = \ang{18},\,\ang{54},\,\ang{198}$ and $\ang{234}$ and for all viewing inclinations. Single-peaked profiles are for $\phi = \ang{342}$ and $\phi = \ang{126}$.

The results of the same model for mass ratio $q = 0.5$ are shown in Figs.\,\ref{Distant_ES12_q05_PA}-\ref{Distant_ES12_q05_TF}. The $\varphi$ and PO are shown in Figs.\,\ref{Distant_ES12_q05_PA} and \ref{Distant_ES12_q05_PO} respectively and both are following the same trend as it was for the case with mass ratio $q = 1$ and both are showing mild asymmetry in profiles. The TF (Fig.\,\ref{Distant_ES12_q05_TF}) is showing asymmetric double-peak or single-peak profiles with the positions of the peaks varying depending on the observed orbital phase of the system. For $q = 0.1$, simulated profiles for $\varphi$, PO and TF are shown in Figs.\,\ref{Distant_ES12_q01_PA}-\ref{Distant_ES12_q01_TF}. The results are very similar as in the previous case with remarkable asymmetry in the profiles.

\textbf{Contact}: The results for this model are shown in Figs.\,\ref{Contact_ES1_PA}-\ref{Contact_ES1_TF}. The $\varphi$-profiles are shown in Fig.\,\ref{Contact_ES1_PA} for different orbital phase of the system. Profiles are very similar as the ones obtained for \textbf{distant} model, but with greater amplitude of maxima/minima. The PO profiles are shown in Fig.\,\ref{Contact_ES1_PO}. For $\phi = \ang{18},\,\ang{198}$ and $\ang{342}$, the profiles are the same as for the single SMBH scenario, while for all the other azimuthal viewing angles, the maximum PO is in the line core. The TF is shown in Fig.\,\ref{Contact_ES1_TF}. Lines are the broadest when viewed for $\phi = \ang{90}$ and $\ang{270}$. The random velocity component in this model that is present in the BLR flattens the line profiles, making it difficult to distinguish between sinle-peaked and double-peaked profiles.

\textbf{Mixed}: Simulations for this model are shown in Figs.\,\ref{Mixed_ES4_PA}-\ref{Mixed_ES4_TF}. Polarization angle is shown in Fig.\,\ref{Mixed_ES4_PA}. The $\varphi$-profiles are double-peaked for $\phi = \ang{18},\,\ang{126}$ and $\ang{198}$. A swing in the $\varphi$-profile in the line core, common for single SMBH, can is when $\phi = \ang{270}$. The PO is overall flat with very mild features in the line core (Fig.\,\ref{Mixed_ES4_PO}). As explained in the Results section, due to finite number of clouds in the simulations we obtain spiky profiles for TF (Fig.\,\ref{Mixed_ES4_TF}). We could expect that unpolarized lines are single-peaked.

\textbf{Spiral}: The results for this model are shown in Figs.\,\ref{Spiral_ES6_PA}-\ref{Spiral_ES6_TF}. The $\varphi$-profiles are shown in Fig.\,\ref{Spiral_ES6_PA}. This model is unique for having double-peaked $\varphi$-profiles when viewed from all azimuthal angles, similar to those found for \textbf{distant} and \textbf{contact} models, but with lower amplitude. The PO is shown in Fig.\,\ref{Spiral_ES6_PO}. It shows profile common for single SMBH scenario for $\phi = \ang{18},\,\ang{126},\,\ang{126}$ and $\ang{342}$, but also those with maximum PO in the core when viewed for all the other azimuthal viewing angles. The TF is shown in Fig.\,\ref{Spiral_ES6_TF}. For intermediate inclination, it shows clear double-peaked profiles, while for nearly face-on viewing inclinations, a single-peaked profile or profiles with peaks very close to each other, can be seen when $\phi = \ang{18},\,\ang{198}$ and $\ang{342}$.

\end{appendix}

\bibliography{bibliography} 
\bibliographystyle{aa} 

\end{document}